\documentclass[aps,prd,12point,twocolumn,nofootinbib,showpacs,superscriptaddress]{revtex4-2}

\usepackage{graphicx}% Include figure files
\usepackage{dcolumn}% Align table columns on decimal point
\usepackage{tabu}
\usepackage{amsmath,amssymb}
\usepackage{float}
\usepackage{natbib}
\usepackage{balance}
\usepackage[normalem]{ulem}
\usepackage{braket}
\usepackage{upgreek}
\usepackage{mathrsfs}
\usepackage{bm}% bold math
\usepackage{lipsum}
\usepackage{comment}
\usepackage{xfrac}
\usepackage[dvipsnames]{xcolor}
\usepackage[title]{appendix}
\usepackage{txfonts}
\usepackage{xcolor}
\usepackage{slashed}
\usepackage[breaklinks=true,pdfborder={0 0 0},bookmarksopen]{hyperref}
\hypersetup{
	colorlinks   = true, %Colours links instead of ugly boxes
	urlcolor     = magenta, %Colour for external hyperlinks
	linkcolor    = blue, %Colour of internal links
	citecolor   = Red %Colour of citations
}

\begin{document}

\title{Dynamics of charged spin-$\frac{1}{2}$ particles in superposed states minimally coupled to electromagnetism in curved spacetime within the WKB approximation}

\author{F. Hammad}
\email{fhammad@ubishops.ca}
\affiliation{Department of Physics \& Astronomy,
Bishop's University,\\
2600 College Street, Sherbrooke, QC, J1M 1Z7, Canada}
\affiliation{Physics Department, Champlain College-Lennoxville,\\
2580 College Street, Sherbrooke, QC, J1M 0C8, Canada}

\author{R. Saadati}
\email{rsaadati@ubishops.ca}
\affiliation{Department of Physics \& Astronomy,
Bishop's University,\\
2600 College Street, Sherbrooke, QC, J1M 1Z7, Canada}

\author{M. Simard}
\email{msimard23@ubishops.ca}
\affiliation{Department of Physics \& Astronomy,
Bishop's University,\\
2600 College Street, Sherbrooke, QC, J1M 1Z7, Canada}

\author{S. Novoa-Cattivelli}
\email{snovoa23@ubishops.ca}
\affiliation{Department of Physics \& Astronomy,
Bishop's University,\\
2600 College Street, Sherbrooke, QC, J1M 1Z7, Canada}
%%%%%%%%%%%%%%%%%%%%%%%%%%%%%%%%%%%%%%%%%%%%%%%%%%%%%%%%%%%%%%%%%%%%%%%%%%%%%%%%%%%%%%%%%%%%
%%%%%%%%%%%%%%%%%%%%%%%%%%%%%%%%%%%%%%%%%%%%%%%%%%%%%%%%%%%%%%%%%%%%%%%%%%%%%%%%%%%%%%%%%%%%
%%%%%%%%%%%%%%%%%%%%%%%%%%%%%%%%%%%%%%%%%%%%%%%%%%%%%%%%%%%%%%%%%%%%%%%%%%%%%%%%%%%%%%%%%%%%
%%%%%%%%%%%%%%%%%%%%%%%%%%%%%%%%%%%%%%%%%%%%%%%%%%%%%%%%%%%%%%%%%%%%%%%%%%%%%%%%%%%%%%%%%%%%

\begin{abstract}
We investigate the curved-spacetime dynamics of charged spin-$\frac{1}{2}$ particles minimally coupled to the electromagnetic field and propagating in superposed states of different masses. For that purpose, we make use of a Wentzel–Kramers–Brillouin approximation of the Dirac equation with a minimal coupling to the Maxwell field in curved spacetime. We obtain the spin dynamics as well as the deviation from geodesic motion of such particles. The problem of having the mass eigenstates experience different proper times, as opposed to the case of neutral particles, is here dealt with by first extracting the second-order differential equation obeyed by each superposition. This strategy proves to be not only powerful, but also very economical. 
\end{abstract}

%\pacs {03.65.Pm, 04.62.+v, 04.20.-q}
%PACS, the Physics and Astronomy Classification Scheme.
\maketitle
\section{Introduction}\label{sec:Intro}
The dynamics of spinning extended bodies in general relativity is described by two sets of equations, commonly known as the MPD equations as they have been first developed by Mathisson \cite{Mathisson}, re-derived in an improved way by Papapetrou \cite{Papapetrou}, before they got extended further by Dixon to include higher moments of the spinning body \cite{Dixon1964,Dixon1970a,Dixon1970b}. For a brief history of the equations see Refs.\,\cite{Dixon2008,Dixon2015}. The equations describe both the spin dynamics and the equation of motion of the spinning body as the latter moves freely in curved spacetime. The first set of equations describing the motion of the spinning body predicts a deviation from geodesic motion of the body caused by the coupling between the spin of the latter and the spacetime's curvature tensor. This coupling and the deviation from geodesic motion is the most important novelty that came out of the equations. The second set of equations describes the dynamics of the body's spin, entailing a spin precession under the influence of gravity.

The MPD equations, as extended by Dixon, allow one to also take into account the possibility of a spinning body that is charged and moving inside an electromagnetic field in curved spacetime. The equations describe then the combined effect of gravity and electrodynamics on the motion of the body. Concerning the momentum dynamics, the extra terms arising on the right-hand side of the equations due to the charge of the body contain the usual Lorentz force as well as a coupling between the spin of the body and the gradient of the electromagnetic field tensor. Concerning the spin dynamics, the extra terms brought by the charge carried by the body are due to a direct coupling between the spin tensor and the electromagnetic field strength tensor.

Later on, it was shown by R\"udiger \cite{Rudiger} and Audretsch \cite{Audretsch} that MPD-like equations could also be obtained for freely propagating spin-$\frac{1}{2}$ particles in curved spacetime by extracting them from a Wentzel–Kramers–Brillouin (WKB)\footnote{Sometimes it is also referred to it in the literature by the Wentzel–Kramers–Brillouin-Jeffreys (WKBJ) approximation \cite{W,K,B,J}.} approximation of the curved-spacetime Dirac equation. Their method, inspired by the works in Refs.\,\cite{Pauli,RubinowKeller,RafanelliSchiller}, was then extended by other authors to include massless spinning particles as well as charged massive spinning particles coupled to the electromagnetic field (see also \cite{Deriglazov1,Deriglazov2,Deriglazov3} and the more recent works \cite{Oancea,Maniccia1,Maniccia2,Maniccia3} and references therein\footnote{See the works \,\cite{Pavsic,Barut,GaioliGarcia,SilenkoTeryaev,NuriUnal,CianfraniMontani1,CianfraniMontani2,ObukhovSilenkoTeryaev} and references therein for other approaches for obtaining spin dynamics of fermions in curved spacetime.}.)

Now, it is well known that quantum particles might exhibit another peculiar feature besides having an intrinsic spin. Indeed, quantum particles could be in a superposition of two different states that might even carry different masses like in the case of neutrinos. As each state carries its own mass and its own spin, searching for MPD-like equations governing such superposed physical quantities in curved spacetime becomes of great importance.

The task of applying the WKB approximation to the curved spacetime Dirac equation for extracting MPD-like equations for neutral particles freely propagating in superposed states of different masses has recently been carried out in Ref.\,\cite{SuperposedPrecession1}. Our goal in this paper is to extend such a study to include charged spin-$\frac{1}{2}$ particles minimally coupled to the electromagnetic field. The main problem that one is confronted with is the fact that the mass eigenstates experience, as opposed to the case of neutral particles, different proper times. As such, taking the proper-time derivative of the superposition of such states to extract the various equations of MPD-like dynamics becomes, indeed, ill-defined. The way out is to first extract from the coupled Dirac equations obeyed by the superpositions a second-order differential equation involving only one of the superpositions at a time. The WKB approximation should then be applied to that second-order differential equation. This strategy proves to be, indeed, very powerful and very economical. A comparison between the results obtained with this method and those obtained in Ref.\,\cite{SuperposedPrecession1} for neutral particles by applying the WKB approximation to the mass eigenstates and their respective equations will be given and discussed at length.   

The rest of this paper is structured as follows. In Sec.\,\ref{sec:MPDIntro}, we briefly recall the two sets of classical MPD equations governing the motion and spin dynamics of extended and charged spinning bodies in curved spacetime in the presence of an electromagnetic field. In Sec.\,\ref{sec:MPDfromWKB}, we apply the WKB approximation to the curved-spacetime Dirac equation minimally coupled to the Maxwell field and derive MPD-like equations for charged spin-$\frac{1}{2}$ particles in a way (different from the one adopted in Ref.\,\cite{Oancea}) that supplies us with the tools to deal with the multi-state particles case. We use those results and tools in Sec.\,\ref{sec:SuperposedMPD} to tackle the case of charged spin-$\frac{1}{2}$ particles propagating in curved spacetime as a superposition of two different quantum states carrying different masses. Our main findings will be briefly summarized in Sec.\,\ref{sec:Conclusion}. The more involved calculations are worked out in detail in appendices \ref{App1} through  \ref{App4}.

%%%%%%%%%%%%%%%%%%%%%%%%%%%%%%%%%%%%%%%%%%%%%%%%%%%%%%%%%%%%%%%%%%%%%%%%%%%%%%%%%%%%%%%%%%%%%%%%%%%%%%%%%%%%%%%%%%%%%%%%
\section{The classical MPD equations for charged spinning bodies}\label{sec:MPDIntro}
The first set of the classical MPD equations describing the equation of motion of a charged spinning body, of mass $m$ and of charge $e$, minimally coupled to electromagnetism is given, at the pole-dipole approximation, by \cite{Dixon1964}\footnote{We adopt, throughout this paper, the spacetime metric signature $(-,+,+,+)$, and we set $c=1$.}:
\begin{equation}\label{MPD1}
\dot{p}^\mu\equiv\frac{{\rm D}p^\mu}{{\rm d}\tau}=-e\varv_\nu F^{\mu\nu}-\tfrac{1}{2}R^\mu_{\;\;\nu\rho\sigma}\varv^\nu S^{\rho\sigma}-\tfrac{1}{2}Q_{\nu\rho}\nabla^\mu F^{\nu\rho}.
\end{equation}
An overdot denotes here, and henceforth, a proper-time derivative; $\tau$ being the proper time. The body's center-of-mass 4-velocity is $\varv^\mu={\rm d}x^\mu/{\rm d}\tau$. The total covariant derivative is denoted by ${\rm D}={\rm d}x^\mu\nabla_\mu$, and $R^\mu_{\;\;\nu\rho\sigma}$ is the Riemann curvature tensor to which the spin tensor $S^{\mu\nu}$ of the spinning body is coupled. The field strength $F_{\mu\nu}$ associated to the Maxwell field $A^\mu$ is $F_{\mu\nu}=\nabla_\mu A_\nu-\nabla_\nu A_\mu$, and the kinematical 4-momentum of the body is $m\varv^\mu=P^\mu+eA^\mu$, where $P_\mu=\partial_\mu\mathcal{S}$ is the generalized 4-momentum extracted from the action functional $\mathcal{S}$ associated to the spinning body. The 4-vector $p^\mu$ appearing in Eq.\,(\ref{MPD1}) represents the dynamical 4-momentum of the body, which is neither identical to the kinematical 4-momentum $m\varv^\mu$ nor necessarily parallel to the latter. The tensor $Q_{\mu\nu}$ represents the electromagnetic moments of the spinning body \cite{Dixon1964,Dixon1970a}. The first term on the right-hand side of Eq.\,(\ref{MPD1}) represents the familiar Lorentz force. Note that even in the absence of the Maxwell field, the right-hand side of Eq.\,(\ref{MPD1}) does not vanish, meaning that the spinning body still deviates from geodesic motion due to its spinning in a spacetime of non-zero curvature.

The spin-angular momentum vector $S^\mu$ of the body is extracted from the spin tensor $S^{\mu\nu}$ via the dynamical 4-momentum $p^\mu$, the mass $m$ and the spacetime metric by $S^\mu=-\frac{1}{2m}\epsilon^\mu_{\;\;\nu\rho\lambda}p^\nu S^{\rho\lambda}$. The spacetime metric enters in the totally antisymmetric Levi-Civita tensor $\epsilon_{\mu\nu\rho\lambda}=\sqrt{-g}\,\varepsilon_{\mu\nu\rho\lambda}$ via the square root of its determinant $g$. The normalization of the Levi-Civita alternating symbol $\varepsilon_{\mu\nu\rho\lambda}$ is $\varepsilon_{0123}=1$. The dynamics of the spin tensor $S^{\mu\nu}$ is described by the second set of MPD equations that read \cite{Dixon1964},
\begin{equation}\label{MPD2}
\dot{S}^{\mu\nu}\equiv\frac{{\rm D}S^{\mu\nu}}{{\rm d}\tau}=p^\mu \varv^\nu-p^\nu \varv^\mu+Q^{\mu\rho}{F_\rho}^{\nu}-Q^{\nu\rho}{F_\rho}^{\mu}.
\end{equation}
Note, again, that even in the absence of the Maxwell field, this equation does not reduce to $\dot{S}^{\mu\nu}=0$, for $\varv^\mu$ and $p^\mu$ might not be parallel to each other.

As Eqs.\,(\ref{MPD1}) and (\ref{MPD2}) are not fully determinate, the supplementary condition involving the spin tensor and the dynamical 4-momentum, $S^{\mu\nu}p_\nu=0$ \cite{Moller,Tulczyjew}, often known as the Tulczyjew-M{\o}ller condition, is imposed by hand in order to amend the system of seven independent equations (\ref{MPD1}) and (\ref{MPD2}) to help solve for the ten unknowns $\varv^\mu$, $\varv_\mu p^\mu$ and $S^{\mu\nu}$ (see also Ref.\,\cite{KyrianSemerak} and references therein.)
%%%%%%%%%%%%%%%%%%%%%%%%%%%%%%%%%%%%%%%%%%%%%%%%%%%%%%%%%%%%%%%%%%%%%%%%%%%%%%%%%%%%%%%%%%%%%%%%%%%%%%%%%%%%%%%%%%%%%%%%
\section{MPD-like equations from the Dirac equation}\label{sec:MPDfromWKB}
As we are interested in charged spin-$\frac{1}{2}$ particles interacting with the electromagnetic field, we need to use the Dirac equation coupled to the Maxwell field. As we shall deal with multi-state particles, to distinguish the different possible states a particle can be in we shall use Latin subscripts $i$ and $j$ for states carrying different masses $m_i$ and $m_j$, respectively. However, no summation is intended here and henceforth whenever such indices are repeatedly displayed. Also, we assume that the different states carry the same charge. For a particle of charge $e$ and of mass $m_i$, described by the spinor field $\Psi_i(x)$, the Dirac equation reads
\begin{equation}\label{DiracEq}
\left(i\hbar\gamma^\mu\nabla_\mu-e\gamma^\mu A_\mu-m_i\right)\Psi_i(x)=0.
\end{equation}
In curved spacetime, one replaces in the Dirac equation the flat-spacetime gamma matrices $\gamma^a$ by the spacetime-dependent gamma matrices $\gamma^\mu=e^\mu_a\gamma^a$, where $e^\mu_a$ are the spacetime vierbeins defined for any spacetime metric $g_{\mu\nu}$ and its inverse $g^{\mu\nu}$ using the Minkowski metric $\eta_{ab}$ by $e^\mu_ae^\nu_b\eta^{ab}=g^{\mu\nu}$. Also, the usual partial derivative $\partial_\mu$ is here replaced by the spin covariant derivative, defined, using the spin connection $\omega^{ab}_\mu$, by $\nabla_\mu=\partial_\mu+\tfrac{1}{8}\omega^{ab}_\mu[\gamma_a,\gamma_b]$, where $\omega_\mu^{\,\,ab}=-e^{\nu b}\partial_\mu e^a_\nu+e^{\nu b}\Gamma_{\mu\nu}^\lambda e^{a}_\lambda$. Here, $e^a_\mu$ are the inverse vierbeins such that $e^a_\mu e^\mu_b=\delta^a_b$, where $\delta^a_b$ is the Kronecker delta, and $\Gamma^\lambda_{\mu\nu}$ are the Christoffel symbols associated to the metric. 

The WKB approximation scheme applied to the Dirac equation consists in expanding the Dirac field $\Psi_i(x)$ according to 
the following ansatz:
\begin{equation}\label{WKBAnsatz}
    \Psi_i(x)=\exp\left[\frac{i}{\hbar}\mathcal{S}_i(x)\right]\sum_{n=0}^\infty\hbar^n\psi_i^{(n)}(x).
\end{equation}
The phase function $\mathcal{S}_i(x)$ in this expansion is real, whereas $\psi_i^{(n)}(x)$ are four-component spinors. It is worth emphasising here two important points about this ansatz. The first is that one needs to assume that both the spinors $\psi_i^{(n)}(x)$ and the rate of change of the phase functions $\mathcal{S}_i(x)$ do not vary significantly over the characteristic lengths of both the gravitational and electromagnetic fields. These fields should also vary slowly over the de Broglie wavelength of the particle in each of its possible states $\Psi_i(x)$. The second point is that the ansatz (\ref{WKBAnsatz}) is incomplete when considering particle-creation environments such as in strong gravitational fields \cite{Hawking} and/or in strong electromagnetic fields \cite{Sauter,HeisenbergEuler,Schwinger}, for one has then to also include the negative-energy states in the WKB expansion \cite{Gill,Piazza,OertellSchutzhold}. However, such more general cases will be considered in more detail in a future work.

By inserting the ansatz (\ref{WKBAnsatz}) into the Dirac equation (\ref{DiracEq}) and equating to zero each coefficient in front of each power of $\hbar$, one finds the following set of equations:
\begin{align}
&\left(\gamma^\mu\pi_{i\mu}+m_i\right)\psi_i^{(0)}=0\label{0thOrder},\\
&\left(\gamma^\mu\pi_{i\mu}+m_i\right)\psi_i^{(n)}=i\gamma^\mu\nabla_\mu\psi_i^{(n-1)},\quad n=1,2,3,...\label{1stOrder}
\end{align}
Here, we set $\pi_{i\mu}=\partial_\mu \mathcal{S}_i+eA_\mu$, which we identify with the kinematical 4-momentum $m_i\varv_i^\mu$ of the particle in the state described by the spinor field $\Psi_i(x)$. Therefore, we also have $\nabla_\mu\pi_{i\nu}-\nabla_\nu\pi_{i\mu}=eF_{\mu\nu}$. From the algebraic equation (\ref{0thOrder}), we conclude that $\det(\gamma^\mu\pi_{i\mu}+m_i)=0$ as this is what guarantees that the equation has a nontrivial solution $\psi_i^{(0)}$. This 
 corresponds, as expected from a 4-momentum $\pi_i^\mu$, to the mass-shell condition $\pi_{i\mu}\pi_i^{\mu}=-m_i^2$, which, in turn, yields the Hamilton-Jacobi equation for the phase function $\mathcal{S}_i(x)$, that reads $(\partial_\mu \mathcal{S}_i+eA_\mu)(\partial^\mu \mathcal{S}_i+eA^\mu)=-m_i^2$.

The general eigenspinor $\psi_i^{(0)}$ that is a solution to Eq.\,(\ref{0thOrder}) can be written as a linear combination of the two independent eigenspinors $\Theta_{iA}(x)$ and $\Theta_{iB}(x)$ of the rank-2 matrix multiplying $\psi_i^{(0)}$ in that equation. Therefore, we write
\begin{equation}\label{IndependentSolutions}
\psi_i^{(0)}(x)=a_i^{(0)}(x)\Theta_{iA}(x)+b_i^{(0)}(x)\Theta_{iB}(x).
\end{equation}
Here, $a_i^{(0)}(x)$ and $b_i^{(0)}(x)$ are some complex scalar factors. Note that, in contrast to the case where the Maxwell field is absent, the 4-spinors $\Theta_{iA}(x)$ and $\Theta_{iB}(x)$ do carry here an index $i$ since even after factoring out the inertial mass $m_i$ in Eq.\,(\ref{0thOrder}) the 4-velocity $\varv_i^\mu$ involved in $\pi_i^\mu=m_i\varv_i^\mu$ still distinguishes between the different states. Choosing these spinors to be orthogonal to each other and normalized, with $\bar{\Theta}_{iA}=\Theta_{iA}^\dagger\gamma^0$ and $\bar{\Theta}_{iB}=\Theta_{iB}^\dagger\gamma^0$ denoting the adjoint spinors, we easily extract the following four equations:
\begin{subequations}
\begin{align}
&\bar{\Theta}_{iA}\Theta_{iB}=\delta_{AB},\quad\bar{\Theta}_{iA}\gamma^\mu\Theta_{iB}=\varv_i^\mu\delta_{AB},\label{ThetaA&B}\\
&\pi_i^\mu\nabla_\mu\Theta_{iA}=C_{i1}\Theta_{iA}+C_{i2}\Theta_{iB}+\xi_{iA},\label{piNablaThetaA&Ba}\\
&\pi_i^\mu\nabla_\mu\Theta_{iB}=D_{i1}\Theta_{iB}+D_{i2}\Theta_{iA}+\xi_{iB}.\label{piNablaThetaA&Bb}
\end{align}
\end{subequations}
The second identity in Eq.\,(\ref{ThetaA&B}) follows at once from the orthogonality condition and from $\Theta_{iA}$ and $\Theta_{iB}$ being solutions to Eq.\,(\ref{0thOrder}). The constraints in Eqs.\,(\ref{piNablaThetaA&Ba}) and (\ref{piNablaThetaA&Bb}) are derived in Appendix \ref{App1}, where the expressions of the two new spinors $\xi_{iA}$ and $\xi_{iB}$ are given explicitly in Eqs.\,(\ref{XiA}) and (\ref{XiB}), respectively. Also,
using the constraints (\ref{ThetaA&B})-(\ref{piNablaThetaA&Bb}), together with the Lorentz-force equation $\pi_i^\mu\nabla_\mu\pi_i^\nu=e\pi_{i\mu} F^{\mu\nu}$, we obtain (see Appendix \ref{App2} for the detailed steps of the derivation):
\begin{align}\label{ThetaGammaNablaTheta}
\bar{\Theta}_{iA}\gamma^\mu\nabla_\mu\Theta_{iA}&=(1/m_i)\left[\tfrac{1}{2}\nabla_\mu \pi_i^\mu+\tfrac{i}{4}(\bar{\Theta}_{iA}\sigma^{\mu\nu}\Theta_{iA})eF_{\mu\nu}+C_{i1}\right],\nonumber\\ 
\bar{\Theta}_{iB}\gamma^\mu\nabla_\mu\Theta_{iB}&=(1/m_i)\left[\tfrac{1}{2}\nabla_\mu \pi_i^\mu+\tfrac{i}{4}(\bar{\Theta}_{iB}\sigma^{\mu\nu}\Theta_{iB})eF_{\mu\nu}+D_{i1}\right],\nonumber\\ 
\bar{\Theta}_{iA}\gamma^\mu\nabla_\mu\Theta_{iB}&=(1/m_i)\left[\tfrac{i}{4}(\bar{\Theta}_{iA}\sigma^{\mu\nu}\Theta_{iB})eF_{\mu\nu}+D_{i2}\right].   
\end{align}
We introduced here the standard notation for the gamma matrices commutator: $\sigma^{\mu\nu}=\tfrac{i}{2}[\gamma^\mu,\gamma^\nu]$.

On the other hand, being nonhomogeneous and linear, Eqs.\,(\ref{1stOrder}) are solvable if and only if the 4-spinors $\bar{\Theta}_{iA}(x)$ and $\bar{\Theta}_{iB}(x)$ are orthogonal to the term $\gamma^\mu\nabla_\mu\psi_i^{(n-1)}$ responsible for the non-homogeneity of the equation; that is,
$\bar{\Theta}_{iA}\gamma^\mu\nabla_\mu\psi_i^{(n-1)}=0$ and $\bar{\Theta}_{iB}\gamma^\mu\nabla_\mu\psi_i^{(n-1)}=0$ for any integer $n\geq1$. For $n=1$, we obtain from these conditions in combination with Eq.\,(\ref{IndependentSolutions}) the following:
\begin{align}\label{piNabla(a&b)}
\pi_i^\mu\nabla_\mu a_i^{(0)}(x)&=-\left[\tfrac{1}{2}\nabla_\mu\pi_i^\mu+\tfrac{i}{4}(\bar{\Theta}_{iA}\sigma^{\mu\nu}\Theta_{iA})eF_{\mu\nu}+C_{i1}\right]a_i^{(0)}(x)\nonumber\\
&\quad-\left[\tfrac{i}{4}(\bar{\Theta}_{iA}\sigma^{\mu\nu}\Theta_{iB}) eF_{\mu\nu}+D_{i2}\right]b_i^{(0)}(x),\nonumber\\
\pi_i^\mu\nabla_\mu b_i^{(0)}(x)&=-\left[\tfrac{1}{2}\nabla_\mu\pi_i^\mu+\tfrac{i}{4}(\bar{\Theta}_{iB}\sigma^{\mu\nu}\Theta_{iB}) eF_{\mu\nu}+D_{i1}\right]b_i^{(0)}(x)\nonumber\\
&\quad-\left[\tfrac{i}{4}(\bar{\Theta}_{iB}\sigma^{\mu\nu}\Theta_{iA})eF_{\mu\nu}+C_{i2}\right]a_i^{(0)}(x).
\end{align}
Note that by setting $F_{\mu\nu}=0$ in these expressions (i.e., in the absence of the Maxwell field), we recover the equations of motion obtained for freely propagating spinning particles \cite{SuperposedPrecession1}. These expressions allow us to extract the following dynamics for the zeroth-order spinor fields $\psi_i^{(0)}(x)$:
\begin{align}\label{piNablapsi(o)}
\pi^\mu_i\nabla_\mu{\psi}_i^{(0)}(x)&=-\tfrac{1}{2}\left(\nabla_\mu \pi_i^\mu\right)\psi_i^{(0)}(x)\nonumber\\
&-\tfrac{i}{4}\left[(\bar{\Theta}_{iA}\sigma^{\mu\nu}\psi^{(0)}_i)\Theta_{iA}+(\bar{\Theta}_{iB}\sigma^{\mu\nu}\psi^{(0)}_i)\Theta_{iB}\right]eF_{\mu\nu}\nonumber\\
&-\tfrac{1}{2m_i}\left[(\bar{\Pi}_{iA}\gamma^{\mu}\psi^{(0)}_i)\Pi_{iA}+(\bar{\Pi}_{iB}\gamma^{\mu}\psi^{(0)}_i)\Pi_{iB}\right]e\pi_i^\nu F_{\mu\nu}.
\end{align}
The orthogonal and normalized 4-spinors $\Pi_{iA}$ and $\Pi_{iB}$ are defined in Appendix \ref{App1}, and use has been made here of identities (\ref{XiA}) and (\ref{XiB}) derived in that appendix. As the 4-spinors $\Pi_{iA}$ and $\Pi_{iB}$ satisfy the equation $(\gamma^\mu\pi_{i\mu}-m_i)\Pi_i=0$, a similar reasoning leading to Eqs.\,(\ref{ThetaA&B})-(\ref{ThetaGammaNablaTheta}) and (\ref{XiA}) and (\ref{XiB}) also gives here the following constraints on such spinors:
\begin{subequations}
\begin{align}\label{PiA&B}
&\bar{\Pi}_{iA}\Pi_{iB}=-\delta_{AB},\quad\bar{\Pi}_{iA}\gamma^\mu\Pi_{iB}=\varv_i^\mu\delta_{AB},\\
&\pi_i^\mu\nabla_\mu\Pi_{iA}=K_{i1}\Pi_{iA}+K_{i2}\Pi_{iB}+\zeta_{iA},\nonumber\\
&\pi_i^\mu\nabla_\mu\Pi_{iB}=L_{i1}\Pi_{iB}+L_{i2}\Pi_{iA}+\zeta_{iB},\label{piNablaPiA&B}
\end{align}
\end{subequations}
where the four complex factors $K_{i1}$, $K_{i2}$, $L_{i1}$ and $L_{i2}$ also satisfy, $K_{i1}=-K^*_{i1}$, $L_{i1}=-L^*_{i1}$ and $K_{i2}=-L^*_{i2}$, the explicit expressions of the 4-spinors $\zeta_{iA}$ and $\zeta_{iB}$ are given by Eqs.\,(\ref{ZetaiA}) and (\ref{ZetaiB}), respectively, and
\begin{align}\label{PiGammaNablaPi}
\bar{\Pi}_{iA}\gamma^\mu\nabla_\mu\Pi_{iA}&=(1/m_i)\left[\tfrac{1}{2}\nabla_\mu \pi_i^\mu-\tfrac{i}{4}(\bar{\Pi}_{iA}\sigma^{\mu\nu}\Pi_{iA})eF_{\mu\nu}+K_{i1}\right],\nonumber\\ 
\bar{\Pi}_{iB}\gamma^\mu\nabla_\mu\Pi_{iB}&=(1/m_i)\left[\tfrac{1}{2}\nabla_\mu \pi_i^\mu-\tfrac{i}{4}(\bar{\Pi}_{iB}\sigma^{\mu\nu}\Pi_{iB})eF_{\mu\nu}+L_{i1}\right],\nonumber\\ 
\bar{\Pi}_{iA}\gamma^\mu\nabla_\mu\Pi_{iB}&=(1/m_i)\left[-\tfrac{i}{4}(\bar{\Pi}_{iA}\sigma^{\mu\nu}\Pi_{iB})eF_{\mu\nu}+L_{i2}\right].   
\end{align}
Using Eq.\,(\ref{piNablapsi(o)}), we obtain the following important identity:
\begin{equation}\label{barpsi0piNablapsi0}
\bar{\psi}_i^{(0)}\pi^\mu_i\nabla_\mu\psi^{(0)}_i=-\tfrac{1}{2}\left(\nabla_\mu \pi_i^\mu\right)\bar{\psi}_i^{(0)}\psi_i^{(0)}-\tfrac{i}{4}(\bar{\psi}_i^{(0)}\sigma^{\mu\nu}\psi^{(0)}_i)eF_{\mu\nu}.
\end{equation}

Now, the minimally coupled Dirac equation (\ref{DiracEq}) yields a conserved Dirac 4-current density $\bar{\Psi}_i\gamma^\mu\Psi_i$, the Gordon decomposition of which reads:
\begin{multline}\label{GordonDecomp}
    j_i^\mu(x)=\frac{i\hbar}{2m_i}\left(\nabla^\mu\bar{\Psi}_i\Psi_i-\bar{\Psi}_i\nabla^\mu\Psi_i\right)+\frac{e}{m_i}A^\mu\bar{\Psi}_i\Psi_i\\
    +\frac{\hbar}{2m_i}\nabla_\nu\left(\bar{\Psi}_i\sigma^{\mu\nu}\Psi_i\right).
\end{multline}
The convection 4-current density $j_{ic}^\mu(x)$ carrying the interaction with the Maxwell field is represented by the first and second terms of the sum. The last term represents the spin 4-current density $j^\mu_{is}(x)$. These currents are both conserved as can easily be checked using the Dirac equation (\ref{DiracEq}). We then define, as in Ref.\,\cite{Rudiger}, the dynamical 4-momentum to be associated to particle $i$ in the state $\Psi_i(x)$ by $p_i^\mu=m_ij_{ic}^{\mu}/(\bar{\Psi}_i\Psi_i)$. Inserting the WKB ansatz (\ref{WKBAnsatz}) into this definition, we get, up to the first order in $\hbar$, the following expression:
\begin{equation}\label{QuantumDynamical4-momentum}
 p_i^\mu(x)=\pi_i^\mu+\frac{i\hbar}{2\bar{\psi}_i^{(0)}\psi_i^{(0)}}\left[\nabla^\mu\bar{\psi}_i^{(0)}\psi_i^{(0)}-\bar{\psi}_i^{(0)}\nabla^\mu\psi_i^{(0)}\right],
\end{equation}
from which it easily follows that (see Appendix \ref{App3}):
\begin{equation}\label{QMPD1}
\pi_i^\mu\nabla_\mu p_i^\nu=e\pi_{i\mu} F^{\mu\nu}-\tfrac{1}{2}R^{\nu}_{\;\;\mu\rho\sigma} \pi_i^\mu S_i^{\rho\sigma}-\tfrac{1}{2}eS_{i\mu\rho}\nabla^\nu F^{\mu\rho}-(\nabla^\nu \pi_{i\mu})p_i^\mu.
%-(\nabla^\nu \pi_{i\mu})S^{\mu\lambda}\pi_i^\rho eF_{\lambda\rho}.  
\end{equation}
The spin tensor $S_i^{\mu\nu}$ appearing in this equation is associated to the particle in the state $\Psi_i(x)$, the explicit definition of which reads \cite{Audretsch,Rudiger}: 
\begin{equation}\label{SpinTensor}
S_i^{\mu\nu}=\hbar\frac{\bar{\Psi}_i\sigma^{\mu\nu}\Psi_i}{2\bar{\Psi}_i{\Psi}_i}=\hbar\frac{\bar{\psi}_i^{(0)}\sigma^{\mu\nu}\psi_i^{(0)}}{2\bar{\psi}_i^{(0)}{\psi}_i^{(0)}}+\mathcal{O}(\hbar^2).
\end{equation}
Only the leading term, which is first-order in $\hbar$, has been kept in the second step. We immediately check using this definition that, at first order in $\hbar$, each state $\Psi_i(x)$ satisfies the Tulczyjew-M{\o}ller condition $ S^{\mu\nu}_ip_{i\nu}=0$. 
Moreover, from the definition (\ref{SpinTensor}) we easily work out the derivative $\pi_i^\rho\nabla_\rho S_i^{\mu\nu}$ by making use of Eq.\,(\ref{piNablapsi(o)}). We find
\begin{equation}\label{QMPD2}
    \pi_i^\rho\nabla_\rho S_i^{\mu\nu}=eS_i^{\mu\rho}F_\rho\,^\nu-eS_i^{\nu\rho}F_\rho\,^\mu.
\end{equation}

Comparing the two results (\ref{QMPD1}) and (\ref{QMPD2}) with the classical MPD equations (\ref{MPD1}) and (\ref{MPD2}), we first notice the appearance of an extra term on the right-hand side of Eq.\,(\ref{QMPD1}) that is not present in Eq.\,(\ref{MPD1}) and the absence in Eq.\,(\ref{QMPD2}) of the first two terms on the right-hand side of Eq.\,(\ref{MPD2}). This pattern is the same as the one found in the case of freely propagating spin-$\frac{1}{2}$ particles in Ref.\,\cite{SuperposedPrecession1}. It was pointed out in the latter work that these two differences between the quantum and classical equations do not seem to arise within a purely Lagrangian approach as reported in Refs.\,\cite{Pavsic,Barut,GaioliGarcia,NuriUnal}. Nevertheless, both equations reduce at the zeroth order in $\hbar$ to the expected classical results; namely, to the geodesic equation $\dot{p}_i^\mu=0$ of a classical point particle for Eq.\,(\ref{QMPD1}) and to $\dot{S}^{\mu\nu}=0$ for Eq.\,(\ref{QMPD2}). 

On the other hand, the remaining terms in Eqs.\,(\ref{QMPD1}) and (\ref{QMPD2}) are identical to those of their classical counterparts (\ref{MPD1}) and (\ref{MPD2}), respectively, provided one identifies the product $eS^{\mu\nu}$ with the magnetic dipole moment $Q^{\mu\nu}$ of the spinning particle times the mass of the latter, as it was first done by Souriau in Ref.\,\cite{Souriau}.

%%%%%%%%%%%%%%%%%%%%%%%%%%%%%%%%%%%%%%%%%%%%%%%%%%%%%%%%%%%%%%%%%%%%%%%%%%%%%%%%%%%%%%%%%%%%%%%%%%%%%%%%%%%%%%%%%%%%%%%%
\section{The dynamical equations for superposed states}\label{sec:SuperposedMPD}
Let a particle $I$ of mass $m_I$ be in a superposition of two different quantum states $\Psi_1(x)$ and $\Psi_2(x)$ carrying masses $m_1$ and $m_2$, respectively. We associate to such a particle a spinor field $\Phi_I(x)$ made of a linear superposition of those two spinors. In doing so, we automatically end up with another particle $II$ of mass $m_{II}$ in a state $\Phi_{II}(x)$ made of a linear combination of $\Psi_1(x)$ and $\Psi_2(x)$ obtained by an orthogonal rotation in the state space $(\Psi_1,\Psi_2)$ as follows: 
\begin{align}\label{PhiI&PhiII}
    \Phi_I(x)&=\Psi_1(x)\cos\theta+
\Psi_2(x)\sin\theta,\nonumber\\
\Phi_{II}(x)&=\Psi_2(x)\cos\theta-\Psi_1(x)
\sin\theta.
\end{align}
In both combinations, we take the parameter $\theta$ to be constant and real. 

In Ref.\,\cite{SuperposedPrecession1}, the dynamics of the particles in the superposed states $\Phi_I(x)$ and $\Phi_{II}(x)$ was found by working out the proper-time derivatives of the momenta and of the spin tensors associated to those superpositions using the combinations (\ref{PhiI&PhiII}) together with the results obtained for the individual spinors $\Psi_1(x)$ and $\Psi_2(x)$ in the previous section. However, such a strategy would not properly work here. Indeed, unlike the free propagation case studied in Ref.\,\cite{SuperposedPrecession1}, the individual states $\Psi_1(x)$ and $\Psi_2(x)$ do neither experience here the same proper time $\tau$ nor acquire the same 4-velocity $\varv^\mu$. This is caused by their coupling to the electromagnetic field that prevents those states from following spacetime geodesics. 

To get around such a difficulty, we shall follow here a different strategy that consists of working exclusively with the superpositions $\Phi_I(x)$ and $\Phi_{II}(x)$ thanks to the coupled Dirac equations they satisfy. The methods and tools developed in the previous section for dealing with the spinors $\Psi_1(x)$ and $\Psi_2(x)$ can then be adopted to deal with the superpositions $\Phi_I(x)$ and $\Phi_{II}(x)$. In fact, from the combinations (\ref{PhiI&PhiII}) it follows that the Dirac equations obeyed by the spinors $\Phi_I(x)$ and $\Phi_{II}(x)$ are coupled and read
\begin{align}\label{CoupledDirac}
\left(i\hbar\gamma^\mu\nabla_\mu-e\gamma^\mu A_\mu-m_{I}\right)\Phi_{I}(x)&=m_{I,II}\Phi_{II}(x),\nonumber\\
\left(i\hbar\gamma^\mu\nabla_\mu-e\gamma^\mu A_\mu-m_{II}\right)\Phi_{II}(x)&=m_{I,II}\Phi_{I}(x).
\end{align}
The individual masses $m_I$ and $m_{II}$ of the two particles and the coupling mass $m_{I,II}$ are given in terms of the masses $m_1$ and $m_2$ associated to the states $\Psi_1(x)$ and $\Psi_2(x)$ by
\begin{align}\label{MassesI&II}
m_{I}&=m_1\cos^2\theta+m_2\sin^2\theta,\quad m_{II}=m_1\sin^2\theta+m_2\cos^2\theta,\nonumber\\
m_{I,II}&=\tfrac{1}{2}\Delta m_{21}\sin2\theta.
\end{align}

In order to apply the WKB approximation scheme to the coupled Dirac equations (\ref{CoupledDirac}), we expand the Dirac fields $\Phi_I(x)$ and $\Phi_{II}(x)$ according to 
the following ans\"atze:
\begin{align}\label{CoupledWKBAnsatzes}
    \Phi_I(x)&=\exp\left[\frac{i}{\hbar}\mathcal{S}_I(x)\right]\sum_{n=0}^\infty\hbar^n\phi_I^{(n)}(x),\nonumber\\
    \Phi_{II}(x)&=\exp\left[\frac{i}{\hbar}\mathcal{S}_{II}(x)\right]\sum_{n=0}^\infty\hbar^n\phi_{II}^{(n)}(x).
\end{align}
The phase functions $\mathcal{S}_I(x)$ and $\mathcal{S}_{II}(x)$ in these expansions are again real, whereas $\phi_I^{(n)}(x)$ and $\phi_{II}^{(n)}(x)$ are four-component spinors. It should also be kept in mind that the remarks concerning the applicability of such WKB expansions made below Eq.\,(\ref{WKBAnsatz}) apply here as well.

It is clear that by inserting these ans\"atze into the Dirac equations (\ref{CoupledDirac}) and equating to zero each coefficient in front of each power of $\hbar$, we would not obtain identities in the form of polynomials in $\hbar$. The way out is to first extract from Eq.\,(\ref{CoupledDirac}) uncoupled second-order differential equations involving only one of the two spinors in each. The result is:
\begin{equation}\label{DecoupledDirac}
\left[\gamma^\mu\gamma^\nu\mathcal{D}_\mu\mathcal{D}_\nu+i(m_I+m_{II})\gamma^\mu\mathcal{D}_\mu-m_{I}m_{II}+m_{I,II}^2\right]\Phi_{I}(x)=0.
\end{equation}
We introduced here, for convenience, the gauge covariant derivative $\mathcal{D}_\mu=\hbar\nabla_\mu+ieA_\mu$. The equation for the spinor $\Phi_{II}$ is obtained by making in Eq.\,(\ref{DecoupledDirac}) the substitution $I\leftrightarrow II$\footnote{It is worth noting here that Eq.\,(\ref{DecoupledDirac}) is, as expected, similar to the equation one obtains by applying twice the Dirac operator to a spinor that does not satisfy the homogeneous Dirac equation \cite{SchroDirac}.}. Before proceeding further with this equation, it is worth pausing here to provide a brief discussion on the latter. Indeed, a few remarks concerning our transition from the first-order coupled Dirac equations (\ref{CoupledDirac}) to the second-order differential equation (\ref{DecoupledDirac}) are in order. 

In fact, Eq.\,(\ref{DecoupledDirac}) requires the knowledge of both the initial value of the spinor field $\Phi_I$ and that of its first derivatives. In contrast, the two equations (\ref{CoupledDirac}), each being a first-order equation, might seem to require only the knowledge of the initial values of the spinor fields. However, as those equations are coupled, solving either one of them for one of the two spinors still requires the knowledge of the first derivatives of the other spinor. As a consequence, combining the two equations (\ref{CoupledDirac}) into the second-order differential equations (\ref{DecoupledDirac}) (one for $\Phi_I$ and the other for $\Phi_{II}$) leads to two independent equations that automatically require the knowledge of the first derivatives of their spinor fields. The spinor field solutions one obtains for Eq.\,(\ref{DecoupledDirac}) are, hence, not redundant as all the information of the interaction between the two spinors is now simply carried by the coupling mass $m_{I,II}$ and by the involvement of the first derivatives among the required initial conditions, all in the same equation. 

In what follows, we introduce, also for convenience, the reduced mass $\mathfrak{m}$ and the total mass $M$ given, respectively, by
\begin{equation}\label{muIM}
\mathfrak{m}=\frac{m_Im_{II}-m_{I,II}^2-\pi_I^2}{m_I+m_{II}},\qquad
M=m_I+m_{II}.    
\end{equation}
We defined here $\pi_{I\mu}=\partial_\mu\mathcal{S}_I+eA_\mu$ to be identified with the kinematical 4-momentum of particle $I$. By inserting the WKB ansatz for $\Phi_I(x)$ from Eq.\,(\ref{CoupledWKBAnsatzes}) into Eq.\,(\ref{DecoupledDirac}), and keeping only terms that are up to the first order in $\hbar$, we obtain the following two equations after separately equating to zero the sum of all the terms multiplying $\hbar^0$ and $\hbar^1$, respectively:
\begin{align}
\left(\gamma^\mu\pi_{I\mu}+\mathfrak{m}\right)\phi_I^{(0)}&=0,\label{0thOrderI}\\
\left(\gamma^\mu\pi_{I\mu}+\mathfrak{m}\right)\phi_I^{(1)}&=\frac{i}{M}\left[\gamma^\mu\gamma^\nu(\nabla_\mu\pi_{I\nu})+M\gamma^\mu\nabla_\mu-2\pi_I^\mu\nabla_{\mu}\right]\phi_I^{(0)}.\label{1stOrderI}
\end{align}
The algebraic equation (\ref{0thOrderI}) implies that $\pi_{I\mu}\pi_I^{\mu}=-\mathfrak{m}^2$, which suggests the definition $\pi_I^\mu=\mathfrak{m}\varv_I^\mu$ for the effective kinematical 4-momentum of particle $I$, with $\varv^\mu_I$ being the 4-velocity associated to its classical trajectory parameterized by a proper time $\tau_I$. Plugging this back into the first identity in Eq.\,(\ref{muIM}), we obtain the two possible solutions $\mathfrak{m}=m_1$ or $\mathfrak{m}=m_2$. For consistency, one has to stick, as it is shown in Appendix \ref{App4}, to the same choice throughout.

Equation (\ref{0thOrderI}) resembles Eq.\,(\ref{0thOrder}). Therefore, using the same reasoning that led us to obtain the explicit expressions of the spinors $\psi_i^{(0)}(x)$, we find the zeroth-order spinor $\phi_I^{(0)}(x)$ to be given in terms of two orthogonal and normalized spinors $\Theta_{IA}(x)$ and $\Theta_{IB}(x)$ as follows:  
\begin{equation}\label{IIndependentSolutions}
\phi_I^{(0)}(x)=a_I^{(0)}(x)\Theta_{IA}(x)+b_I^{(0)}(x)\Theta_{IB}(x),
\end{equation}
where $a_I^{(0)}(x)$ and $b_I^{(0)}(x)$ are complex coefficients, and $\Theta_{IA}(x)$ and $\Theta_{IB}(x)$ are the two linearly independent solutions of the algebraic equation (\ref{0thOrderI}). The latter spinors also obey the following constrains:  
\begin{subequations}
\begin{align}\label{ThetaIA&IB}
&\bar{\Theta}_{IA}\Theta_{IB}=\delta_{AB},\quad\bar{\Theta}_{IA}\gamma^\mu\Theta_{IB}=\varv^\mu_I\delta_{AB},\\
&\pi_I^\mu\nabla_\mu\Theta_{IA}=C_{I1}\Theta_{IA}+C_{I2}\Theta_{IB}+\xi_{IA},\label{piNablaThetaIA&IBa}\\
&\pi_I^\mu\nabla_\mu\Theta_{IB}=D_{I1}\Theta_{IB}+D_{I2}\Theta_{IA}+\xi_{IB}.\label{piNablaThetaIA&IBb}
\end{align}
\end{subequations}
The expressions of the two new spinors $\xi_{IA}$ and $\xi_{IB}$ are identical to those of $\xi_{iA}$ and $\xi_{iB}$ as given explicitly in Eqs.\,(\ref{XiA}) and (\ref{XiB}), respectively, provided only that one replaces everywhere the index $i$ by $I$. Similarly, the Lorentz-force equation $\pi_I^\mu\nabla_\mu\pi_I^\nu=e\pi_{I\mu} F^{\mu\nu}$ yields the following additional identities:
\begin{align}\label{SuperposedThetaGammaNablaTheta}
\bar{\Theta}_{IA}\gamma^\mu\nabla_\mu\Theta_{IA}&=(1/\mathfrak{m})\left[\tfrac{1}{2}\nabla_\mu \pi_I^\mu+\tfrac{i}{4}(\bar{\Theta}_{IA}\sigma^{\mu\nu}\Theta_{IA})eF_{\mu\nu}+C_{I1}\right],\nonumber\\ 
\bar{\Theta}_{IB}\gamma^\mu\nabla_\mu\Theta_{IB}&=(1/\mathfrak{m})\left[\tfrac{1}{2}\nabla_\mu \pi_I^\mu+\tfrac{i}{4}(\bar{\Theta}_{IB}\sigma^{\mu\nu}\Theta_{IB})eF_{\mu\nu}+D_{I1}\right],\nonumber\\ 
\bar{\Theta}_{IA}\gamma^\mu\nabla_\mu\Theta_{IB}&=(1/\mathfrak{m})\left[\tfrac{i}{4}(\bar{\Theta}_{IA}\sigma^{\mu\nu}\Theta_{IB})eF_{\mu\nu}+D_{I2}\right]. 
\end{align}

On the other hand, in contrast to Eqs.\,(\ref{1stOrder}), the solvability of the nonhomogeneous linear equation (\ref{1stOrderI}) requires the following extra constraints to be satisfied:
\begin{align}\label{ConstraintsonI,II}
\bar{\Theta}_{IA}\left[\gamma^\mu\gamma^\nu(\nabla_\mu\pi_{I\nu})+M\gamma^\mu\nabla_\mu-2\pi_I^\mu\nabla_{\mu}\right]\phi_I^{(0)}&=0,\nonumber\\
\bar{\Theta}_{IB}\left[\gamma^\mu\gamma^\nu(\nabla_\mu\pi_{I\nu})+M\gamma^\mu\nabla_\mu-2\pi_I^\mu\nabla_{\mu}\right]\phi_I^{(0)}&=0.
\end{align}
Plugging the decomposition 
(\ref{IIndependentSolutions}) into these constraints and taking account of Eqs.\,(\ref{ThetaIA&IB}) and (\ref{SuperposedThetaGammaNablaTheta}), we obtain the dynamics of the coefficients $a^{(0)}_I(x)$ and $b^{(0)}_{I}(x)$ to be:
\begin{align}\label{piNabla(aI&bI)}
\pi_I^\mu\nabla_\mu a_I^{(0)}&=-\left[\tfrac{1}{2}\nabla_\mu\pi_I^\mu+\tfrac{i}{4}(\bar{\Theta}_{IA}\sigma^{\mu\nu}\Theta_{IA})eF_{\mu\nu}+C_{I1}\right]a_I^{(0)}\nonumber\\
&\quad-\left[\tfrac{i}{4}(\bar{\Theta}_{IA}\sigma^{\mu\nu}\Theta_{IB}) eF_{\mu\nu}+D_{I2}\right]b_I^{(0)},\nonumber\\
\pi_I^\mu\nabla_\mu b_I^{(0)}&=-\left[\tfrac{1}{2}\nabla_\mu\pi_I^\mu+\tfrac{i}{4}(\bar{\Theta}_{IB}\sigma^{\mu\nu}\Theta_{IB}) eF_{\mu\nu}+D_{I1}\right]b_I^{(0)}\nonumber\\
&\quad-\left[\tfrac{i}{4}(\bar{\Theta}_{IB}\sigma^{\mu\nu}\Theta_{IA})eF_{\mu\nu}+C_{I2}\right]a_I^{(0)}.
\end{align}
Remarkably, these identities are, \textit{mutatis mutandis}, the same as in Eq.\,(\ref{piNabla(a&b)}). Using these, together with Eqs.\,(\ref{ThetaIA&IB})-(\ref{piNablaThetaIA&IBb}), we extract the following dynamics for the zeroth-order spinor field $\phi_I^{(0)}(x)$:
\begin{align}\label{piNablapsiI(o)}
\pi^\mu_I\nabla_\mu{\phi}_I^{(0)}(x)&=-\tfrac{1}{2}\left(\nabla_\mu \pi_I^\mu\right)\phi_I^{(0)}(x)\nonumber\\
&-\tfrac{i}{4}\left[(\bar{\Theta}_{IA}\sigma^{\mu\nu}\phi^{(0)}_I)\Theta_{IA}+(\bar{\Theta}_{IB}\sigma^{\mu\nu}\phi^{(0)}_I)\Theta_{IB}\right]eF_{\mu\nu}\nonumber\\
&-\tfrac{1}{2\mathfrak{m}}\left[(\bar{\Pi}_{IA}\gamma^{\mu}\phi^{(0)}_I)\Pi_{IA}+(\bar{\Pi}_{IB}\gamma^{\mu}\phi^{(0)}_I)\Pi_{IB}\right]e\pi_I^\nu F_{\mu\nu}.
\end{align}
Combining this result with Eqs.\,(\ref{ThetaIA&IB}) and (\ref{ConstraintsonI,II}), we obtain the following two additional results, respectively:
\begin{align}\label{barpsiI0piNablapsiI0}
\bar{\phi}_I^{(0)}\pi^\mu_I\nabla_\mu\phi^{(0)}_I&=-\tfrac{1}{2}\left(\nabla_\mu \pi_I^\mu\right)\bar{\phi}_I^{(0)}\phi_I^{(0)}-\tfrac{i}{4}(\bar{\phi}_I^{(0)}\sigma^{\mu\nu}\phi^{(0)}_I)eF_{\mu\nu},\\
\bar{\phi}_I^{(0)}\gamma^\mu\nabla_\mu\phi^{(0)}_I&=0.\label{barpsiI0gammaNablapsiI0}
\end{align}
All these equations are identical to those obtained for single-state particles. The difference will now manifest itself in the dynamical equations satisfied by the 4-momenta and the spin tensors to be associated to each of the two superposed states $\Phi_I$ and $\Phi_{II}$. 

In Ref.\,\cite{SuperposedPrecession1}, it was found that there are three options for building the dynamical 4-momentum to be associated to each superposition. The first option consists in replacing the spinors $\Psi_1$ and $\Psi_2$ by the spinors $\Phi_I$ and $\Phi_{II}$ in the expressions (\ref{QuantumDynamical4-momentum}) giving the momenta $p_1^\mu$ and $p_2^\mu$ of the single-state particles. The result is one pair of 4-momenta denoted, respectively, by $p_I^{A\mu}$ and $p_{II}^{A\mu}$. The second option consists of extracting a 4-momentum from the Gordon decomposition of the non-conserved current densities $\bar{\Phi}_I\gamma^\mu\Phi_I$ and $\bar{\Phi}_{II}\gamma^\mu\Phi_{II}$. The result is a second pair of 4-momenta denoted by $p_I^{B\mu}$ and $p_{II}^{B\mu}$. The last option is based on extracting a single 4-momentum from the Gordon decomposition of the conserved total current density $\bar{\Phi}_I\gamma^\mu\Phi_I+\bar{\Phi}_{II}\gamma^\mu\Phi_{II}$. The result is a `mixed'
4-momentum denoted by $p_{I,II}^\mu$. In what follows, we are going to study the dynamics of each one of these 4-momenta separately before discussing the physical meaning of each and their differences. 
%%%%%%%%%%%%%%%%%%%%%%%%%%
%%%%%% Subsection 1 %%%%%%
%%%%%%%%%%%%%%%%%%%%%%%%%%
\subsection{Dynamics of $p_I^{A\mu}$}
As discussed in the introduction, all our dynamical equations can now naturally be obtained with respect to the proper time of each superposition of interest rather than with respect to a unique proper time $\tau$ that would be common to the individual mass eigenstates. This is made possible thanks to working directly with the superpositions $\Phi_I$ and $\Phi_{II}$ rather than with their components $\Psi_1$ and $\Psi_2$. In what follows, all the dynamical equations will be derived with respect to the proper time $\tau_I$ associated to the superposition $\Phi_I$. The simple index substitution $I\rightarrow II$ is what returns the corresponding equations for the superposition $\Phi_{II}$.   

To first order in $\hbar$, the explicit expression of the 4-momentum $p_I^{A\mu}$ obtained from the first option is
\begin{align}\label{pIA}
 p_I^{A\mu}(x)&=\frac{i\hbar}{2\bar{\Phi}_I\Phi_I}\left(\nabla^\mu\bar{\Phi}_I\Phi_I-\bar{\Phi}_I\nabla^\mu\Phi_I\right)+eA^\mu\nonumber\\
 &=\pi_I^{\mu}+\frac{i\hbar}{2\bar{\phi}_I^{(0)}\phi_I^{(0)}}\left[\nabla^\mu\bar{\phi}_I^{(0)}\phi_I^{(0)}-\bar{\phi}_I^{(0)}\nabla^\mu\phi_I^{(0)}\right].
\end{align}
An identical calculation leading to the result (\ref{QMPD1}) then gives the following dynamical equation with respect to the proper time $\tau_I$:
\begin{equation}\label{SuperposedQMPD1A}
\pi_I^\mu\nabla_\mu p_I^{A\nu}=e\pi_{I\mu} F^{\mu\nu}-\tfrac{1}{2}R^{\nu}_{\;\;\mu\rho\sigma} \pi_I^\mu S_I^{\rho\sigma}-\tfrac{1}{2}eS_{I\mu\rho}\nabla^\nu F^{\mu\rho}-(\nabla^\nu \pi_{I\mu})p_I^{A\mu}. 
\end{equation}
The spin tensor $S_I^{\mu\nu}$ is here defined, in analogy to Eq.\,(\ref{SpinTensor}), by
\begin{equation}\label{SI}
S_I^{\mu\nu}=\hbar\frac{\bar{\Phi}_I\sigma^{\mu\nu}\Phi_I}{2\bar{\Phi}_I{\Phi}_I}=\hbar\frac{\bar{\phi}_I^{(0)}\sigma^{\mu\nu}\phi_I^{(0)}}{2\bar{\phi}_I^{(0)}{\phi}_I^{(0)}}+\mathcal{O}(\hbar^2).
\end{equation}
Using this expression, it is straightforward to check that at first order in $\hbar$ the Tulczyjew-M{\o}ller condition takes here the form $S_I^{\mu\nu}p_{I\nu}^A=0$. It is also easy to check, using Eqs.\,(\ref{pIA}), (\ref{App4PhiIIPhiII}) and (\ref{App4NablaPhiIIPhiII}), that at first order in $\hbar$ we have $p_I^{A\mu}=p_{II}^{A\mu}$, whereas using the coupled Dirac equations (\ref{CoupledDirac}) we learn that $S_I^{\mu\nu}=S_{II}^{\mu\nu}$ at first order in $\hbar$. The two particles display the same spin tensor and acquire, only within this first option as we shall see shortly, the same dynamical 4-momentum as well.
%%%%%%%%%%%%%%%%%%%%%%%%%%
%%%%%% Subsection 2 %%%%%%
%%%%%%%%%%%%%%%%%%%%%%%%%%
\subsection{Dynamics of $p_I^{B\mu}$}
Using the coupled Dirac equations (\ref{CoupledDirac}), we easily derive the following decomposition of the current density $\bar{\Phi}_I\gamma^\mu\Phi_I$:
%%%%%%%%%%%%%%%
\begin{align}\label{GordonPhiIGammaPhiI}
\bar{\Phi}_I\gamma^\mu\Phi_I&=\frac{i\hbar  m_{II}}{2(m_Im_{II}-m^2_{I,II})}\left[\nabla^\mu\bar{\Phi}_I\Phi_I-\bar{\Phi}_I\nabla^\mu\Phi_I\right.\nonumber\\
&\quad\left.-\frac{m_{I,II}}{m_{II}}(\nabla^\mu\bar{\Phi}_{II}\Phi_{I}-\bar{\Phi}_{I}\nabla^\mu\Phi_{II})\right]\nonumber\\
&\quad+\frac{m_{II}eA_\mu}{m_Im_{II}-m^2_{I,II}}\left[\bar{\Phi}_I\Phi_I-\frac{m_{I,II}}{2m_{II}}\left(\bar{\Phi}_I\Phi_{II}+\bar{\Phi}_{II}\Phi_I\right)\right]\nonumber\\
&\quad+\frac{\hbar m_{II}}{2(m_Im_{II}-m^2_{I,II})}\left[\nabla_\nu(\bar{\Phi}_I\sigma^{\mu\nu}\Phi_I)\right.\nonumber\\
&\quad\left.-\frac{m_{I,II}}{m_{II}}(\nabla_\nu\bar{\Phi}_{II}\sigma^{\mu\nu}\Phi_{I}+\bar{\Phi}_{I}\sigma^{\mu\nu}\nabla_\nu\Phi_{II})\right]\nonumber\\
&\quad+\frac{im_{I,II}eA_\nu}{2(m_Im_{II}-m^2_{I,II})}\left(\bar{\Phi}_{II}\sigma^{\mu\nu}\Phi_I-\bar{\Phi}_I\sigma^{\mu\nu}\Phi_{II}\right).
\end{align}
The current density $\bar{\Phi}_{II}\gamma^\mu\Phi_{II}$ yields a decomposition that is identical to Eq.\,(\ref{GordonPhiIGammaPhiI}) modulo the index substitution $I\leftrightarrow II$. 

Despite not being conserved,
the terms in the first three lines on the right-hand side of Eq.\,(\ref{GordonPhiIGammaPhiI}) can be viewed as making a convection 4-current density $J_{Ic}^\mu(x)$. The terms in the remaining lines can be viewed as making a spin current density $J_{Is}^{\mu}(x)$ that is not conserved either. We can use the current $J_{Ic}^\mu(x)$ to build a 4-momentum $p_I^{B\mu}(x)$ that we can associate to the first particle by setting $p_I^{B\mu}=m_IJ_{Ic}^\mu/(\bar{\Phi}_I\Phi_I)$. Therefore, to first order in $\hbar$, the explicit expression of such a 4-momentum reads, after using the results (\ref{App4PhiIIPhiI}) and (\ref{App4NablaPhiIIPhiI}):
\begin{align}\label{pIB}
 p_I^{B\mu}(x)&=\frac{i\hbar m_I m_{II}}{m_Im_{II}-m^2_{I,II}}\left(\frac{\nabla^\mu\bar{\Phi}_I\Phi_I-\bar{\Phi}_I\nabla^\mu\Phi_I}{2\bar{\Phi}_I\Phi_I}\right.\nonumber\\
 &\quad\left.-\frac{m_{I,II}}{m_{II}}\frac{\nabla^\mu\bar{\Phi}_{II}\Phi_{I}-\bar{\Phi}_{I}\nabla^\mu\Phi_{II}}{2\bar{\Phi}_I\Phi_I}\right)\nonumber\\
&\quad+\frac{m_I m_{II}eA^\mu}{m_Im_{II}-m^2_{I,II}}\left(1-\frac{m_{I,II}}{2m_{II}}\frac{\bar{\Phi}_I\Phi_{II}+\bar{\Phi}_{II}\Phi_I}{\bar{\Phi}_I\Phi_I}\right)\nonumber\\
&=\frac{m_I(M-\mathfrak{m})}{m_Im_{II}-m^2_{I,II}}p_I^{A\mu}(x).
\end{align}
It is worth noting here that, in contrast to the momenta $p_I^{A\mu}$ and $p_{II}^{A\mu}$, the two momenta $p_I^{B\mu}$ and $p_{II}^{B\mu}$ are not identical to each other.
Using now the result (\ref{SuperposedQMPD1A}), identity (\ref{pIB}) immediately gives the proper-time evolution of the 4-momentum $p_I^{B\mu}$ to be:
\begin{align}\label{SuperposedQMPD1B}
&\pi_I^\mu\nabla_\mu p_I^{B\nu}=\nonumber\\
&\quad\frac{m_I(M-\mathfrak{m})}{m_Im_{II}-m^2_{I,II}}\left(e\pi_{I\mu} F^{\mu\nu}-\tfrac{1}{2}R^{\nu}_{\;\;\mu\rho\sigma} \pi_I^\mu S_I^{\rho\sigma}-\tfrac{1}{2}eS_{I\mu\rho}\nabla^\nu F^{\mu\rho}\right)\nonumber\\
&\quad-(\nabla^\nu \pi_{I\mu})p_I^{B\mu}.
\end{align}
%%%%%%%%%%%%%%%%%%%%%%%%%%
%%%%%% Subsection 3 %%%%%%
%%%%%%%%%%%%%%%%%%%%%%%%%%
\subsection{Dynamics of $p_{I,II}^{\mu}$}
Using the expression (\ref{GordonPhiIGammaPhiI}) of the current density $\bar{\Phi}_I\gamma^\mu\Phi_I$, we arrive at the following sum:
\begin{align}\label{PhiIPhiI+PhiIIPhiII}
&\bar{\Phi}_I\gamma^\mu\Phi_I+\bar{\Phi}_{II}\gamma^\mu\Phi_{II}\nonumber\\
 &=\frac{m_I m_{II}}{(m_Im_{II}-m^2_{I,II})}\left[\frac{i\hbar}{2m_I}\left(\nabla^\mu\bar{\Phi}_I\Phi_I-\bar{\Phi}_I\nabla^\mu\Phi_I\right)\right.\nonumber\\
 &\quad\left.+\frac{i\hbar}{2m_{II}}\left(\nabla^\mu\bar{\Phi}_{II}\Phi_{II}-\bar{\Phi}_{II}\nabla^\mu\Phi_{II}\right)\right]\nonumber\\
 &\quad-\frac{i\hbar m_{I,II}\left(\nabla^\mu\bar{\Phi}_{II}\Phi_{I}-\bar{\Phi}_{I}\nabla^\mu\Phi_{II}+\nabla^\mu\bar{\Phi}_{I}\Phi_{II}-\bar{\Phi}_{II}\nabla^\mu\Phi_{I}\right)}{2(m_Im_{II}-m^2_{I,II})}\nonumber\\
 &\quad+\frac{m_I m_{II}eA^\mu}{m_Im_{II}-m^2_{I,II}}\left(\frac{\bar{\Phi}_I\Phi_I}{m_I}\!+\!\frac{\bar{\Phi}_{II}\Phi_{II}}{m_{II}}\!-\!\frac{m_{I,II}}{m_Im_{II}}\left[\bar{\Phi}_I\Phi_{II}\!+\!\bar{\Phi}_{II}\Phi_I\right]\right)\nonumber\\
 &\quad+\frac{m_I m_{II}}{2(m_Im_{II}-m^2_{I,II})}\nabla_\nu\left(\frac{\hbar}{m_I}[\bar{\Phi}_I\sigma^{\mu\nu}\Phi_I]+\frac{\hbar}{m_{II}}[\bar{\Phi}_{II}\sigma^{\mu\nu}\Phi_{II}]\right)\nonumber\\
 &\quad-\frac{\hbar m_{I,II}}{2(m_Im_{II}-m^2_{I,II})}\nabla_\nu\left(\bar{\Phi}_{II}\sigma^{\mu\nu}\Phi_{I}+\bar{\Phi}_{I}\sigma^{\mu\nu}\Phi_{II}\right).
\end{align}
The terms in the first four lines on the right-hand side of this identity can be viewed as making a conserved convection 4-current density $J_{cI,II}^\mu(x)$. The terms in the remaining lines on the right-hand side can be viewed as making a spin current density $J_{sI,II}^{\mu}(x)$ that is also conserved. We can then use the current $J_{cI,II}^\mu(x)$ to build a 4-momentum $p_{I,II}^\mu(x)$ that we can associate to the mixed states $\Phi_I$ and $\Phi_{II}$ by setting, as is done in Ref.\,\cite{SuperposedPrecession1}: $p_{I,II}^{\mu}=(m_Im_{II}-m_{I,II}^2)^{1/2}J_{cI,II}^\mu/(\bar{\Sigma}\Sigma)$, where we defined $\bar{\Sigma}\Sigma=\bar{\Phi}_I\Phi_I+\bar{\Phi}_{II}\Phi_{II}$.

Indeed, this definition of $p_{I,II}^\mu$ is suggested by the fact that we can write the currents $J^\mu_{cI,II}(x)$ and $J^\mu_{s,I,II}(x)$ extracted from Eq.\,(\ref{PhiIPhiI+PhiIIPhiII}) in a more condensed way as:
\begin{align}\label{CondensedConvection&Spin}
J_{cI,II}^\mu(x)&=\tfrac{i}{2}\hbar\left(\nabla^\mu\bar{\Sigma}{\mathbb M}^{-1}\Sigma-\bar{\Sigma}{\mathbb M}^{-1}\nabla^\mu\Sigma\right)+eA^\mu\bar{\Sigma}\mathbb{M}^{-1}\Sigma,\nonumber\\
J_{sI,II}^\mu(x)&=\tfrac{1}{2}\hbar\nabla_\nu\left(\bar{\Sigma}\Gamma^{\mu\nu}{\mathbb M}^{-1}\Sigma\right),    
\end{align}
where we introduced for convenience the following notations:
\begin{equation}
\Gamma^\mu=\begin{pmatrix}
            \gamma^\mu & 0\\
            0 & \gamma^\mu
           \end{pmatrix},\quad {\mathbb M}=\begin{pmatrix}
               m_I & m_{I,II}\\
               m_{I,II} & m_{II}
           \end{pmatrix},\quad 
           \Sigma=\begin{pmatrix}
               \Phi_I\\
               \Phi_{II}
           \end{pmatrix}.    
    \end{equation}
Thus, the matrices $\Gamma^\mu$ and $\mathbb M$ in Eq.\,(\ref{CondensedConvection&Spin}) formally act as 2$\times$2 matrices on a 2-component  spinor $\Sigma$, the adjoint of which is taken to be 
$\bar{\Sigma}\equiv\Sigma^{\dagger}\Gamma^0$. Also, in analogy to the matrix $\sigma^{\mu\nu}$, the block-diagonal matrix $\Gamma^{\mu\nu}$ is defined by $\Gamma^{\mu\nu}=\frac{i}{2}[\Gamma^\mu,\Gamma^\nu]$ and has as its nonvanishing elements only $\sigma^{\mu\nu}$ sitting on its two diagonal entries. The expression of $J^\mu_{cI,II}(x)$ in Eq.\,(\ref{CondensedConvection&Spin}) is what suggests the definition $p_{I,II}^\mu=\sqrt{\det({\mathbb M})} J_{cI,II}^\mu/(\bar{\Sigma}\Sigma)$ we chose for the 4-momentum.

Therefore, to first order in $\hbar$, the explicit expression of such a 4-momentum reads, after using again the results (\ref{App4PhiIIPhiI}) and (\ref{App4NablaPhiIIPhiI}):
\begin{align}\label{PI,II}
&p_{I,II}^\mu(x)=\frac{m_I m_{II}}{\sqrt{m_1m_2}}\left(\frac{i\hbar}{2m_I}\frac{\nabla^\mu\bar{\Phi}_I\Phi_I-\bar{\Phi}_I\nabla^\mu\Phi_I}{\bar{\Sigma}\Sigma}\right.\nonumber\\
&\left.+\frac{i\hbar}{2m_{II}}\frac{\nabla^\mu\bar{\Phi}_{II}\Phi_{II}-\bar{\Phi}_{II}\nabla^\mu\Phi_{II}}{\bar{\Sigma}\Sigma}\right)\nonumber\\
&-\frac{i\hbar m_{I,II}}{\sqrt{m_1m_2}}\left(\frac{\nabla^\mu\bar{\Phi}_{II}\Phi_{I}-\bar{\Phi}_{I}\nabla^\mu\Phi_{II}+\nabla^\mu\bar{\Phi}_{I}\Phi_{II}-\bar{\Phi}_{II}\nabla^\mu\Phi_{I}}{2\bar{\Sigma}\Sigma}\right)\nonumber\\
&+\frac{m_I m_{II}}{\sqrt{m_1m_2}}eA^\mu\left(\frac{\bar{\Phi}_I\Phi_I}{m_I\bar{\Sigma}\Sigma}+\frac{\bar{\Phi}_{II}\Phi_{II}}{m_{II}\bar{\Sigma}\Sigma}-\frac{m_{I,II}}{m_Im_{II}}\frac{\bar{\Phi}_I\Phi_{II}+\bar{\Phi}_{II}\Phi_I}{\bar{\Sigma}\Sigma}\right)\nonumber\\
%%%%%%%%%%%%%%%%%%%%%%%%%%%%%%%%%%%%%%%%%%%
&=\frac{M-\mathfrak{m}}{\sqrt{m_1m_2}}\left[\frac{\bar{\Phi}_I\Phi_I}{\bar{\Sigma}\Sigma}p_I^{A\mu}(x)+\frac{\bar{\Phi}_{II}\Phi_{II}}{\bar{\Sigma}\Sigma}p_{II}^{A\mu}(x)\right]=\frac{M-\mathfrak{m}}{\sqrt{m_1m_2}}p_I^{A\mu}(x).
\end{align}
Note that this result displays, as it should, an explicit symmetry with respect to the substitution $I\leftrightarrow II$.  Using this last expression together with Eq.\,(\ref{SuperposedQMPD1A}), we arrive at the following dynamical equation:
\begin{align}\label{SuperposedQMPD1I,II}
\pi_I^\mu\nabla_\mu p_{I,II}^{\nu}&=\frac{M-\mathfrak{m}}{\sqrt{m_1m_2}}\left(e\pi_{I\mu} F^{\mu\nu}-\tfrac{1}{2}R^{\nu}_{\;\;\mu\rho\sigma} \pi_I^\mu S_I^{\rho\sigma}\!-\!\tfrac{1}{2}eS_{I\mu\rho}\nabla^\nu F^{\mu\rho}\right)\nonumber\\
&\quad-(\nabla^\nu \pi_{I\mu})p_{I,II}^{\mu}.
\end{align}
%%%%%%%%%%%%%%%%%%%%%%%%%%
%%%%%% Subsection 4 %%%%%%
%%%%%%%%%%%%%%%%%%%%%%%%%%
\subsection{Dynamics of $S_I^{\mu\nu}$, $S_{II}^{\mu\nu}$ and $S_{I,II}^{\mu\nu}$}
Taking the proper-time derivative of the spin tensor $S_I^{\mu\nu}$ as given by the definition (\ref{SI}), we find
\begin{align}\label{piNablaSI}
    \pi_I^\mu\nabla_\mu S_I^{\rho\lambda}&=\hbar\frac{\pi_I^\mu\nabla_\mu\bar{\phi}_I^{(0)}\sigma^{\rho\lambda}\phi_I^{(0)}}{2\bar{\phi}_I^{(0)}{\phi}_I^{(0)}}+\hbar\frac{\bar{\phi}_I^{(0)}\sigma^{\rho\lambda}\pi_I^\mu\nabla_\mu\phi_I^{(0)}}{2\bar{\phi}_I^{(0)}{\phi}_I^{(0)}}\nonumber\\
    &\quad-S_I^{\rho\lambda}\frac{\pi_I^\mu\nabla_\mu(\bar{\phi}_I^{(0)}\phi_I^{(0)})}{\bar{\phi}_I^{(0)}\phi_I^{(0)}}+\mathcal{O}(\hbar^2)\nonumber\\
&=eS^{\rho\mu}_IF_\mu\,^\lambda-eS^{\lambda\mu}_IF_\mu\,^\rho.
\end{align}
The dynamics equation $\pi_{II}^\mu\nabla_\mu S_{II}^{\rho\lambda}$ of the spin tensor $S_{II}^{\mu\nu}$ associated to the second superposition follows from the result (\ref{piNablaSI}) by performing the substitution $I\leftrightarrow II$.

The mixed spin tensor $S_{I,II}^{\mu\nu}$ is obtained from the spin current $J_{sI,II}^\mu(x)$. Its explicit expression in terms of the spinors $\Phi_I$ and $\Phi_{II}$ and the spin tensors $S_I^{\mu\nu}$ and $S_{II}^{\mu\nu}$ reads,
\begin{align}\label{MixedSpin}
S_{I,II}^{\mu\nu}&=\frac{m_{II}\bar{\Phi}_I\Phi_I}{\sqrt{m_1m_2}\bar{\Sigma}\Sigma}S_I^{\mu\nu}+\frac{m_{I}\bar{\Phi}_{II}\Phi_{II}}{\sqrt{m_1m_2}\bar{\Sigma}\Sigma}S_{II}^{\mu\nu}\nonumber\\
&\quad-\frac{\hbar m_{I,II}(\bar{\Phi}_{II}\sigma^{\mu\nu}\Phi_I+\bar{\Phi}_I\sigma^{\mu\nu}\Phi_{II})}{2\sqrt{m_1m_2}\bar{\Sigma}\Sigma}=\frac{M-\frak{m}}{\sqrt{m_1m_2}}S_I^{\mu\nu}.
\end{align} 
Therefore, using the result (\ref{piNablaSI}), the dynamics of this spin tensor reads
\begin{align}
    \pi_I^\mu\nabla_\mu S_{I,II}^{\rho\lambda}&=\frac{M-\frak{m}}{\sqrt{m_1m_2}}\left(eS^{\rho\mu}_IF_\mu\,^\lambda-eS^{\lambda\mu}_IF_\mu\,^\rho\right).
\end{align}
Given that the spin tensor $S_I^{\mu\nu}$ satisfies at first order in $\hbar$ the Tulczyjew-M{\o}ller condition, we conclude that the tensor $S_{I,II}^{\mu\nu}$ also satisfies $S_{I,II}^{\mu\nu}p_{r\nu}^s=0$ for $r=I,II$ and for $s=A,B$, and that $S_{I,II}^{\mu\nu}p_{I,II\nu}=0$. 
%%%%%%%%%%%%%%%%%%%%%%%%%%%%%%%%%%%%%%%%%%%%%%%%%%%%%%%%%%%%%%%%%%%%%%%%%%%%%%%%%%%%%%%%%%%%%%%%%%%%%%%%%%%%%%%%%%%%%%%%
%%%%%%%%%%%%%%%%%%%%%%%%%%%%%%%%%%%%%%%%%%%%%%%%%%%%%%%%%%%%%%%%%%%%%%%%%%%%%%%%%%%%%%%%%%%%%%%%%%%%%%%%%%%%%%%%%%%%%%%%
\section{Summary and discussion}\label{sec:Conclusion}
We studied in this paper the dynamics of superposed states minimally coupled to electromagnetism in curved spacetime. We first used a WKB approximation to extract from the Dirac equation of charged spin-$\frac{1}{2}$ particles the quantum analog of the classical MPD equations for charged and extended spinning bodies in curved spacetime. We then generalized those results to the case of superposed states $\Phi_I(x)$ and $\Phi_{II}(x)$ made of a linear combination of distinct states $\Psi_I(x)$ and $\Psi_{II}(x)$ carrying each the same charge, but different masses. We included in our analysis all three options introduced in Ref.\,\cite{SuperposedPrecession1} for building the dynamical 4-momentum to be associated to such superpositions and all three associated spin tensors.

Those three options for building the dynamical 4-momentum should, as it was emphasized in Ref.\,\cite{SuperposedPrecession1}, be chosen according to one's needs. If one wishes to study only one of the two superpositions $\Phi_I$ or $\Phi_{II}$, the first option for the 4-momentum as given by  Eq.\,(\ref{pIA}) is what one would want to use. That definition focuses on the dynamics arising from merely having a spinor that carries a charge and a mass, regardless of whether the spinor is made of two interfering ones. In case one is interested instead in the currents $\bar{\Phi}_I\gamma^\mu\Phi_I$ and $\bar{\Phi}_{II}\gamma^\mu\Phi_{II}$, which are non-conserved and associated to each superposition individually, then one would want to rely on the second option (\ref{pIB}) for building the dynamical 4-momentum. For this option, one relies solely on the existence of a Dirac 4-current associated to a spinor, regardless of whether the 4-current is just one part of a total 4-current made of two separate interacting ones. However, in case one is specifically interested in that total 4-current $\bar{\Phi}_I\gamma^\mu\Phi_I+\bar{\Phi}_{II}\gamma^\mu\Phi_{II}$ that is conserved, one has to consider the 4-momentum (\ref{PI,II}) that can be viewed as a `mixed' dynamical 4-momentum.

In order to put each one of these three different options for building a dynamical 4-momentum into a more physical perspective, we invoke here a useful physical observable that can be extracted from each one of them. Let us start with the first option $p_I^{A\mu}$ that is based on Eq.\,(\ref{pIA}). Recall that the expectation value of the convection current given by the right-hand side of the first line in Eq.\,(\ref{pIA}) yields the mass times the classical \textit{group} velocity of $\Phi_I$ if the latter is a \textit{wave packet} formed of positive-energy solutions (see, e.g., Ref.\,\cite{BjorkenDrell}). However, as the right-hand side of Eq.\,(\ref{pIA}) is introduced by simple analogy with expression (\ref{GordonDecomp}) for single-state particles, the group velocity thus obtained is that of any wave packet $\Phi_I$, irrespective of whether the latter is interacting with $\Phi_{II}$. This is the reason why the two momenta $p_I{\vphantom{p_I}}^{A\mu}$ and $p_{II}{\vphantom{p_I}}^{A\mu}$ came out to be identical.

Let us consider now the second option $p_I{\vphantom{p_I}}^{B\mu}$ that is based on expression (\ref{pIB}). The latter is derived from the terms in the first three lines on the right-hand side of Eq.\,(\ref{GordonPhiIGammaPhiI}) that give the convection part of the current $\bar{\Phi}_I\gamma^\mu\Phi_I$. The expectation value of those terms yields not only the classical group velocity of $\Phi_I$ if the latter is a wave packet formed of positive-energy solutions, but involves also interaction terms coming from the cross terms between $\Phi_I$, $\Phi_{II}$ and their derivatives. Those extra terms will introduce an oscillation due to the different phases of the components of the two wave packets. This is very reminiscent of the well-known \textit{Zitterbewegung} one obtains for the classical group velocity when allowing solutions of both signs of energy in a single wave packet \cite{BjorkenDrell}. The oscillation one will obtain here comes from the interference between the two superpositions. This does, indeed, show up in the 4-momentum $p_I{\vphantom{p_I}}^{B\mu}$ by the latter being a weighted version of the 4-momentum $p_I{\vphantom{p_I}}^{A\mu}$ with a weight that also depends on the mass of the second superposition as well as on the coupling mass. 

Consider now the third option $p_{I,II}^{\mu}$ that is based on expression (\ref{PI,II}). The latter is extracted from the convection part of the conserved total current (\ref{PhiIPhiI+PhiIIPhiII}). The classical group velocity one obtains by averaging such a convection current is a weighted combination of the group velocities of the wave packets $\Phi_I$ and $\Phi_{II}$ plus extra interaction terms coming again from the cross terms between $\Phi_I$, $\Phi_{II}$ and their derivatives. An oscillation due to the different phases of the components of the two wave packets will again emerge. However, the oscillation one obtains here will come out more symmetrical with respect to the exchange of the indices $I$ and $II$ as both wave packets are taken into account symmetrically. This, indeed, does also show up in the 4-momentum $p_{I,II}^{\mu}$ by the latter being a symmetric weighted version of the 4-momentum $p_I{\vphantom{p_I}}^{A\mu}$ with respect to the mass of each superposition. These claims, together with the addition of negative-energy states in the case of pair-creation environments as mentioned below Eq.\,(\ref{WKBAnsatz}), will be made more precise and will be studied in more detail in a future work.

Similarly, the spin tensor $S_{I,II}^{\mu\nu}$ can be viewed as a mixed spin tensor associated to a combination of interacting spinors $\Phi_I$ and $\Phi_{II}$, whereas the spin tensors $S_I^{\mu\nu}$ and $S_{II}^{\mu\nu}$ are associated to each of the individual spinors $\Phi_I$ and $\Phi_{II}$ regardless of whether the latter are interacting or not.  

Remarkably, all our dynamical equations took here a much simpler form than those found in Ref.\,\cite{SuperposedPrecession1}. The reason is that the strategy pursued here consists of first extracting the second-order differential equation obeyed by each superposition from the coupled first-order Dirac equations they obeyed. We then applied the WKB approximation to such an equation by expanding the superpositions $\Phi_I(x)$ and $\Phi_{II}(x)$ themselves in powers of $\hbar$ rather than relying on the WKB expansion of the states $\Psi_1(x)$ and $\Psi_2(x)$ individually as it was done in Ref.\,\cite{SuperposedPrecession1}. This approach is dictated by the unavailability of a common proper time to both mass eigenstates due to their interaction with the electromagnetic field and their different masses. This fact allows one, indeed, to derive the dynamics of each superposition by taking proper-time derivatives only with respect to the proper time of the superposition of interest. As a consequence, all our dynamical equations are expressed here in terms of the momenta and the spin tensors associated to the superpositions $\Phi_I(x)$ and $\Phi_{II}(x)$ rather than in terms of the momenta and spin tensors associated to the individual states $\Psi_1(x)$ and $\Psi_2(x)$. We gather here all the dynamical equations expressed, for convenience, solely in terms of the proper-time derivative ---\,denoted by an overdot \,--- with respect to the proper time $\tau_I$ of the first superposition:
\begin{align}
    \dot{p}_I^{A\mu}&=e\varv_{I\nu} F^{\nu\mu}\!-\!\tfrac{1}{2}R^{\mu}_{\;\;\nu\rho\sigma} \varv_I^\nu S_I^{\rho\sigma}\!-\tfrac{1}{2}(e/\mathfrak{m})S_{I\nu\rho}\nabla^\mu F^{\nu\rho}\!-(\nabla^\mu \varv_{I\nu})p_I^{A\nu},\nonumber\\
    \dot{p}_I^{B\mu}&=\!\!\frac{m_I(M-\mathfrak{m})}{m_Im_{II}\!-\!m^2_{I,II}}\!\left[e\varv_{I\nu} F^{\nu\mu}\!-\!\tfrac{1}{2}R^{\mu}_{\;\;\nu\rho\sigma} \varv_I^\nu S_I^{\rho\sigma}\!\!-\!\tfrac{1}{2}(e/\mathfrak{m})S_{I\nu\rho}\!\nabla^\mu F^{\nu\rho}\right]\nonumber\\
&\quad-(\nabla^\mu v_{I\nu})p_I^{B\nu},\nonumber\\
\dot{p}_{I,II}^{\mu}&=\frac{M-\mathfrak{m}}{\sqrt{m_1m_2}}\left[e\varv_{I\nu} F^{\nu\mu}-\tfrac{1}{2}R^{\mu}_{\;\;\nu\rho\sigma} \varv_I^\nu S_I^{\rho\sigma}-\tfrac{1}{2}(e/\mathfrak{m})S_{I\nu\rho}\nabla^\mu F^{\nu\rho}\right]\nonumber\\
&\quad-(\nabla^\mu \varv_{I\nu})p_{I,II}^{\nu},\nonumber\\
\dot{ S}_I^{\mu\nu}&=(e/\mathfrak{m})\left(S^{\mu\rho}_IF_\rho\,^\nu-S^{\nu\rho}_IF_\rho\,^\mu\right),\nonumber\\
\dot{S}_{I,II}^{\mu\nu}&=\frac{e(M-\frak{m})}{\mathfrak{m}\sqrt{m_1m_2}}\left(S^{\mu\rho}_IF_\rho\,^\nu-S^{\nu\rho}_IF_\rho\,^\mu\right).
\end{align}

It is worth noting in this regard that the dynamical equations obtained in this work in terms of the momenta and spin tensors associated to the superpositions $\Phi_I(x)$ and $\Phi_{II}(x)$ would not reduce to those obtained in Ref.\,\cite{SuperposedPrecession1} even in the absence of the electromagnetic field. The fundamental reason is that there is no simple relation between the phase functions $\mathcal{S}_r(x)$ emerging from the WKB expansion of the superpositions $\Phi_r(x)$ for $r=I,II$ and the phase functions $\mathcal{S}_r(x)$ appearing in the WKB expansion of the states $\Psi_r(x)$ for $r=1,2$. As a consequence, the kinematical 4-momenta $\pi_r^{\mu}=\partial^\mu\mathcal{S}_r+eA^\mu$ for $r=I,II$ are neither identical to the kinematical 4-momenta for $r=1,2$ nor can they be expressed in terms of the latter. Therefore, in contrast to the dynamical equations obtained in Ref.\,\cite{SuperposedPrecession1}, our present results cannot be recast in a meaningful way in terms of the 4-momenta $p_1^\mu(x)$ and $p_2^\mu(x)$. Nevertheless, the terms conveying the interference between the mass eigenstates $\Psi_1(x)$ and $\Psi_2(x)$ can still be recovered simply by inserting the linear combinations (\ref{PhiI&PhiII}) into our various results. Actually, the composed nature of the states already manifests itself in the various dynamical equations through the ubiquitous multiplicative factor $M-\frak{m}/\sqrt{m_1m_2}$.
%%%%%%%%%%%%%%%%%%%%%%%%%%%%%%%%%
%%%%%%%%%%%%%%%%%%%%%%%%%%%%%%%%%
%%%%%%%%%%%%%%%%%%%%%%%%%%%%%%%%%
%%%%%%%%%%%%%%%%%%%%%%%%%%%%%%%%%%
%%%%%%%%%%%%%%%%%%%%%%%%%%%%%%%%%%
%%%%%%%%%%%%%%%%%%%%%%%%%%%%%%%%%%
\section*{Acknowledgments}
The authors are grateful to the anonymous referee for
the constructive comments that helped improve our
manuscript. This work is supported by the Natural Sciences and Engineering Research Council of Canada (NSERC) Discovery Grant No. RGPIN-2017-05388, and by the Fonds de Recherche du Québec - Nature et Technologies (FRQNT).

%%%%%%%%%%%%%%%%%%%%%%%%%%%%%%%%%%%%%%%%%%%%%%%%%%%%%%%%%%%%%%%%%%%%%%%%%%%%%%%%%%%%%%%%%%%%%%%%%%%%%%%%%%%%%%%%%%%%%%%%
\appendix
%%%%%%%%%%%%%%%%%%%%%%%%%%%
%%%%%%%%%%%%%%%%%%%%%%%%%%%
%%%% Appendix 1 %%%%%%%%%%%
%%%%%%%%%%%%%%%%%%%%%%%%%%%
%%%%%%%%%%%%%%%%%%%%%%%%%%%
\section{Derivation of the constraints (\ref{piNablaThetaA&Ba}) and (\ref{piNablaThetaA&Bb})}\label{App1}
We derive the constraints (\ref{piNablaThetaA&Ba}) and (\ref{piNablaThetaA&Bb}) as follows. First, since $\Theta_{iA}(x)$ and $\Theta_{iB}(x)$ are solutions to Eq.\,(\ref{0thOrder}), we have $(\gamma^\mu\pi_{i\mu}+m_i)\Theta_{iA}=0$ and $(\gamma^\mu\pi_{i\mu}+m_i)\Theta_{iB}=0$. Therefore, applying the operator $\pi_i^\nu\nabla_\nu$ to the left-hand side of these two identities and taking account of the Lorentz-force equation $\pi_i^\mu\nabla_\mu\pi_i^\nu=e\pi_{i\mu} F^{\mu\nu}$, leads to
\begin{subequations}
    \begin{align}
    (\gamma^\mu\pi_{i\mu}+m_i)(\pi_i^\nu\nabla_\nu\Theta_{iA})&=e\gamma^\mu\pi_i^\nu F_{\mu\nu}\Theta_{iA},\label{Lorentz1a}\\
     (\gamma^\mu\pi_{i\mu}+m_i)(\pi_i^\nu\nabla_\nu\Theta_{iB})&=e\gamma^\mu\pi_i^\nu F_{\mu\nu}\Theta_{iB}.\label{Lorentz1b}
\end{align}
\end{subequations}
These two identities show that the spinors $\pi_i^\nu\nabla_\nu\Theta_{iA}$ and $\pi_i^\nu\nabla_\nu\Theta_{iB}$ are solutions to a nonhomogeneous equation, the homogeneous part of which is simply Eq.\,(\ref{0thOrder}). Therefore, $\pi_i^\nu\nabla_\nu\Theta_{iA}$ and $\pi_i^\nu\nabla_\nu\Theta_{iB}$ can be written as linear combinations of the form,
\begin{subequations}
\begin{align}
\pi_i^\nu\nabla_\nu\Theta_{iA}&=C_{i1}\Theta_{iA}+C_{i2}\Theta_{iB}+\xi_{iA},\label{piNablaThetaA}\\
\pi_i^\nu\nabla_\nu\Theta_{iB}&=D_{i1}\Theta_{iB}+D_{i2}\Theta_{iA}+\xi_{iB}.\label{piNablaThetaB}
\end{align}
\end{subequations}
The four scalars $C_{i1}$, $C_{i2}$, $D_{i1}$ and $D_{i2}$ are arbitrary complex factors, and $\xi_{iA}(x)$ and $\xi_{iB}(x)$ are particular 4-spinor solutions that are both orthogonal to $\Theta_{iA}$ and $\Theta_{iB}$. 

By applying the operator $\pi_i^\mu\nabla_\mu$ to the normalization conditions $\bar{\Theta}_{iA}\Theta_{iA}=\bar{\Theta}_{iB}\Theta_{iB}=1$ and then using the second identity (\ref{piNablaThetaA&Bb}), we learn that $C_{i1}^*=-C_{i1}$ and $D_{i1}^*=-D_{i1}$. By applying the operator $\pi_i^\mu\nabla_\mu$ to the orthogonality condition $\bar{\Theta}_{iA}\Theta_{iB}=0$, we find $D_{i2}=-C_{i2}^*$. On the other hand, let $\Pi_{iA}$ and $\Pi_{iB}$ be the two orthogonal and independent 4-spinor solutions of the homogeneous algebraic equation $(\gamma^\mu\pi_{i\mu}-m_i)\Pi_i=0$. We then have $\bar{\Theta}_{ir}\Pi_{is}=0$ for any $r,s=A,B$. Hence, we can express the 4-spinors $\xi_{iA}$ and $\xi_{iB}$ in terms of $\Pi_{iA}$ and $\Pi_{iB}$ as follows: 
\begin{align}
\xi_{iA}(x)&=E_1\Pi_{iA}(x)+E_2\Pi_{iB}(x),\label{ZetaExpression}\\
\xi_{iB}(x)&=F_1\Pi_{iA}(x)+F_2\Pi_{iB}(x).\label{XiExpression}
\end{align}
We determine $E_1$, $E_2$, $F_1$ and $F_2$ by inserting expressions (\ref{ZetaExpression}) and (\ref{XiExpression}) into the right-hand side of Eqs.\,(\ref{piNablaThetaA}) and (\ref{piNablaThetaB}), and then plugging the result into Eqs.\,(\ref{Lorentz1a}) and (\ref{Lorentz1b}), respectively. As a consequence, we find that $\xi_{iA}$ and $\xi_{iB}$ take the following explicit expressions:
\begin{align}
    \xi_{iA}&=-\frac{e}{2m_i}\left(\bar{\Pi}_{iA}\gamma^\mu \pi_i^\nu F_{\mu\nu}\Theta_{iA}\right)\Pi_{iA}-\frac{e}{2m_i}\left(\bar{\Pi}_{iB}\gamma^\mu\pi_i^\nu F_{\mu\nu}\Theta_{iA}\right)\Pi_{iB}\label{XiA},\\
    \xi_{iB}&=-\frac{e}{2m_i}\left(\bar{\Pi}_{iA}\gamma^\mu\pi_i^\nu F_{\mu\nu}\Theta_{iB}\right)\Pi_{iA}-\frac{e}{2m_i}\left(\bar{\Pi}_{iB}\gamma^\mu\pi_i^\nu F_{\mu\nu}\Theta_{iB}\right)\Pi_{iB}\label{XiB}.
\end{align}
An identical reasoning leads to the following explicit expressions of the spinors $\zeta_{iA}$ and $\zeta_{iB}$ arising in the constraints (\ref{piNablaPiA&B}):
\begin{align}
    \zeta_{iA}&=-\frac{e}{2m_i}\left(\bar{\Theta}_{iA}\gamma^\mu\pi_i^\nu F_{\mu\nu}\Pi_{iA}\right)\Theta_{iA}-\frac{e}{2m_i}\left(\bar{\Theta}_{iB}\gamma^\mu\pi_i^\nu F_{\mu\nu}\Pi_{iA}\right)\Theta_{iB},\label{ZetaiA}\\
    \zeta_{iB}&=-\frac{e}{2m_i}\left(\bar{\Theta}_{iA}\gamma^\mu\pi_i^\nu F_{\mu\nu}\Pi_{iB}\right)\Theta_{iA}-\frac{e}{2m_i}\left(\bar{\Theta}_{iB}\gamma^\mu\pi_i^\nu F_{\mu\nu}\Pi_{iB}\right)\Theta_{iB}.\label{ZetaiB}
\end{align}
%%%%%%%%%%%%%%%%%%%%%%%%%%
%%%%%%%%%%%%%%%%%%%%%%%%%%
%%%% Appendix 2 %%%%%%%%%% 
%%%%%%%%%%%%%%%%%%%%%%%%%%
%%%%%%%%%%%%%%%%%%%%%%%%%%
%%%%%%%%%%%%%%%%%%%%%%%%%%
\section{Derivation of Eq.\,(\ref{ThetaGammaNablaTheta})}\label{App2}
We derive the constraints (\ref{ThetaGammaNablaTheta}) as follows:
\begin{align}
\bar{\Theta}_{iA}\gamma^\mu\nabla_\mu\Theta_{iA}&=-\frac{1}{m_i}\bar{\Theta}_{iA}\gamma^\mu\nabla_\mu(\gamma^\nu \pi_
{i\nu}\Theta_{iA})\nonumber\\
&=-\frac{1}{m_i}\bar{\Theta}_{iA}\gamma^\mu\gamma^\nu(\nabla_\mu \pi_{i\nu})\Theta_{iA}-\frac{1}{m_i}\bar{\Theta}_{iA}\gamma^\mu\gamma^\nu \pi_{i\nu}\nabla_\mu\Theta_{iA}\nonumber\\
&=\frac{1}{m_i}\nabla_\mu \pi_i^\mu+\frac{ie}{2m_i}\bar{\Theta}_{iA}\sigma^{\mu\nu}F_{\mu\nu}\Theta_{iA}\nonumber\\
&\qquad\qquad\;\,\quad\quad+\frac{2}{m_i}\bar{\Theta}_{iA} \pi_i^\mu\nabla_\mu\Theta_{iA}-\bar{\Theta}_{iA} \gamma^\mu\nabla_\mu\Theta_{iA}\nonumber\\
&=\frac{1}{m_i}\left[\tfrac{1}{2}\nabla_\mu \pi_i^\mu+\tfrac{i}{4}(\bar{\Theta}_{iA}\sigma^{\mu\nu}\Theta_{iA})eF_{\mu\nu}+C_{i1}\right].\label{App2ThetaA&B(B.1)}\\
\bar{\Theta}_{iA}\gamma^\mu\nabla_\mu\Theta_{iB}&=\frac{1}{m_i}\left[\tfrac{i}{4}(\bar{\Theta}_{iA}\sigma^{\mu\nu}\Theta_{iB})eF_{\mu\nu}+D_{i2}\right].\label{App2ThetaA&B(B.2)}
\end{align}
Use has been made here of $\gamma^\mu\gamma^\nu=-2g^{\mu\nu}-\gamma^\nu\gamma^\mu$ and $\gamma^\mu\gamma^\nu=-g^{\mu\nu}-i\sigma^{\mu\nu}$, as well as the normalization (\ref{ThetaA&B}) together with the constraints (\ref{piNablaThetaA&Ba}) and (\ref{piNablaThetaA&Bb}). Equation (\ref{App2ThetaA&B(B.2)}) is obtained by following the same steps leading to Eq.\,(\ref{App2ThetaA&B(B.1)}) after making use of the orthogonality relation in Eq.\,(\ref{ThetaA&B}). The expression of the product $\bar{\Theta}_{iB}\gamma^\mu\nabla_\mu\Theta_{iB}$ is obtained from Eq.\,(\ref{App2ThetaA&B(B.1)}) by the replacements $A\rightarrow B$ and $C_{i1}\rightarrow D_{i1}$. 
%%%%%%%%%%%%%%%%%%%%%%%%%%
%%%%%%%%%%%%%%%%%%%%%%%%%%
%%%%%%%%%%%%%%%%%%%%%%%%%% 
%%%%%%%%%%%%%%%%%%%%%%%%%%
%%%%%%%%%%%%%%%%%%%%%%%%%%
%%%%%%%%%%%%%%%%%%%%%%%%%%
%%%%%%%%%%%%%%%%%%%%%%%%%%
%%%%%%%%%%%%%%%%%%%%%%%%%%
%%%%%%%%%%%%%%%%%%%%%%%%%%
%%%% Appendix 3 %%%%%%%%%% 
%%%%%%%%%%%%%%%%%%%%%%%%%%
%%%%%%%%%%%%%%%%%%%%%%%%%%
%%%%%%%%%%%%%%%%%%%%%%%%%%

\section{Derivation of Eq.\,(\ref{QMPD1})}\label{App3}
Taking the proper-time derivative of the dynamical 4-momentum $p_i^\mu$ as given by expression (\ref{QuantumDynamical4-momentum}), and making use of Eq.\,(\ref{barpsi0piNablapsi0}), we find, up to the first order in $\hbar$:
\begin{align}\label{App3FirstStep}
&\pi^\mu_i\nabla_\mu p_i^\nu=e\pi_{i\mu} F^{\mu\nu}+i\hbar\pi^\mu_i\nabla_\mu\left[\frac{\nabla^\nu\bar{\psi}_i^{(0)}\psi_i^{(0)}-\bar{\psi}_i^{(0)}\nabla^\nu\psi_i^{(0)}}{2\bar{\psi}_i^{(0)}\psi_i^{(0)}}\right]\nonumber\\
    &=e\pi_{i\mu} F^{\mu\nu}\!+\!(\nabla_{\mu}\pi^\mu_i)(p_i^\nu\!-\!\pi_i^\nu)\!+\!i\hbar \pi_{i\mu} \frac{\nabla^\mu\nabla^\nu\bar{\psi}_i^{(0)}\psi_i^{(0)}\!-\!\bar{\psi}_i^{(0)}\nabla^\mu\nabla^\nu\psi_i^{(0)}}{2\bar{\psi}_i^{(0)}\psi_i^{(0)}}\nonumber\\
    &\quad+i\hbar\pi_{i\mu} \frac{\nabla^{\nu}\bar{\psi}_i^{(0)}\nabla^{\mu}\psi_i^{(0)}-\nabla^{\mu}\bar{\psi}_i^{(0)}\nabla^{\nu}\psi_i^{(0)}}{2\bar{\psi}_i^{(0)}\psi_i^{(0)}}.
\end{align}

Next, by switching the order of the covariant derivatives in the numerator of the third term in the second line on the right-hand side of Eq.\,(\ref{App3FirstStep}), and using the fact that (see e.g., Ref.\,\cite{SchroDirac}) $[\nabla_\mu,\nabla_\nu]\psi_i^{(0)}=\frac{i}{4}R_{\mu\nu ab}\sigma^{ab}\psi_i^{(0)}$ and $[\nabla_\mu,\nabla_\nu]\bar{\psi}_i^{(0)}=-\frac{i}{4}R_{\mu\nu ab}\bar{\psi}_i^{(0)}\sigma^{ab}$, together with the definition (\ref{SpinTensor}) of the spin tensor $S_i^{\mu\nu}$ to first order in $\hbar$, Eq.\,(\ref{App3FirstStep}) takes the following form:
\begin{align}\label{App3SecondStep}
    \pi^\mu_i\nabla_\mu p_i^\nu&=e\pi_{i\mu} F^{\mu\nu}+(\nabla_\mu \pi_i^\mu)(p_i^\nu-\pi_i^\nu)-\tfrac{1}{2}R^{\nu}_{\;\;\mu\rho\lambda} \pi_i^\mu S_i^{\rho\lambda}\nonumber\\
    &\quad+i\hbar \frac{\nabla^\nu\left[\pi_{i\mu}\nabla^\mu\bar{\psi}_i^{(0)}\psi_i^{(0)}-\bar{\psi}_i^{(0)}\pi_{i\mu} \nabla^\mu{\psi}_i^{(0)}\right]}{2\bar{\psi}_i^{(0)}\psi_i^{(0)}}\nonumber\\
&\quad-(\nabla^\nu \pi_{i\mu})p_i^\mu+i\hbar\pi_{i\mu}\frac{\nabla^\nu\bar{\psi}_i^{(0)} \nabla^\mu\psi_i^{(0)}-\nabla^\mu\bar{\psi}_i^{(0)}\nabla^\nu\psi_i^{(0)}}{\bar{\psi}_i^{(0)}\psi_i^{(0)}}\nonumber\\
&= e\pi_{i\mu} F^{\mu\nu}-\tfrac{1}{2}R^{\nu}_{\;\;\mu\rho\lambda} \pi_i^\mu S_i^{\rho\lambda}-\tfrac{1}{2}eS_{\mu\rho}\nabla^\nu F^{\mu\rho}-(\nabla^\nu \pi_{i\mu})p_i^\mu.
    \end{align}
In the last line, we have used Eq.\,(\ref{piNablapsi(o)}), the identity $
\nabla^\mu\bar{\Theta}_{iA}\Pi_{iB}=-(\bar{\Theta}_{iA}\gamma^\nu\Pi_{iB})\nabla^\mu\pi_{i\nu}/(2m_i)$, which is easily obtained from the defining equation of $\Theta_{iA}$, as well as the completeness relation $\sum\limits_{s=A,B}(\Theta_{is}\bar{\Theta}_{is}-\Pi_{is}\bar{\Pi}_{is})=1$.
%%%%%%%%%%%%%%%%%%%%%%%%%%
%%%%%%%%%%%%%%%%%%%%%%%%%%
%%%%% Appendix 4  %%%%%%%%
%%%%%%%%%%%%%%%%%%%%%%%%%%
\section{Expressions of $\bar{\Phi}_{II}\Phi_I$ and $\hbar\nabla^\mu\bar{\Phi}_{II}\Phi_I$}\label{App4}
To arrive at Eqs.\,(\ref{pIB}) and (\ref{PI,II}), we need the expressions of the products $\bar{\Phi}_{II}\Phi_I$ and $\hbar\nabla^\mu\bar{\Phi}_{II}\Phi_I$. To avoid dealing with products of the two spinors $\Phi_I$ and $\Phi_{II}$, we express the spinor $\Phi_{II}$ in terms of the spinor $\Phi_I$ using Eq.\,(\ref{CoupledDirac}). Therefore, to first order in $\hbar$, we find after using also the WKB ansatz (\ref{WKBAnsatz}) for $\Phi_I$, the following results:
\begin{align}\label{App4PhiIIPhiI}
\bar{\Phi}_{II}\Phi_I&=\frac{1}{m_{I,II}}\left[-i\hbar\nabla_\mu\bar{\Phi}_I\gamma^\mu\Phi_I-eA_\mu\bar{\Phi}_I\gamma^\mu\Phi_I-m_I\bar{\Phi}_I\Phi_I\right]\nonumber\\
    &=\frac{1}{m_{I,II}}\left[-\pi_{I\mu}(\bar{\phi}_I^{(0)}\gamma^\mu\phi_I^{(0)}+\hbar\bar{\phi}_I^{(1)}\gamma^\mu\phi_I^{(0)}+\hbar\bar{\phi}_I^{(0)}\gamma^\mu\phi_I^{(1)})\right.\nonumber\\
    &\quad\left.-i\hbar\nabla_\mu\bar{\phi}_I^{(0)}\gamma^\mu\phi_I^{(0)}-m_I(\bar{\phi}_I^{(0)}\phi_I^{(0)}+\hbar\bar{\phi}_I^{(1)}\phi_I^{(0)}+\hbar\bar{\phi}_I^{(0)}\phi_I^{(1)})\right]\nonumber\\
    &\quad+\mathcal{O}(\hbar^2)=\frac{\mathfrak{m}-m_I}{m_{I,II}}(\bar{\Phi}_I\Phi_I)_{\mathcal{O}(\hbar)}+\mathcal{O}(\hbar^2),
\end{align}
\begin{align}\label{App4PhiIIPhiII}
&\bar{\Phi}_{II}\Phi_{II}=\frac{1}{m_{I,II}}\left[-i\hbar\nabla_\mu\bar{\Phi}_I\gamma^\mu-eA_\mu\bar{\Phi}_I\gamma^\mu-m_I\bar{\Phi}_I\right]\nonumber\\
&\qquad\qquad\times\frac{1}{m_{I,II}}\left[i\hbar\nabla_\mu\Phi_I\gamma^\mu-eA_\mu\Phi_I\gamma^\mu-m_I\Phi_I\right]\nonumber\\
    &=\frac{1}{m_{I,II}^2}\left[-\pi_{I\mu}(\bar{\phi}_{I}^{(0)}+\hbar\bar{\phi}_I^{(1)})\gamma^\mu-i\hbar\nabla_\mu\bar{\phi}_I^{(0)}\gamma^\mu-m_I(\bar{\phi}_I^{(0)}+\hbar\bar{\phi}_I^{(1)})\right]\nonumber\\
&\quad\quad\;\times\left[-\pi_{I\mu}\gamma^\mu(\phi_{I}^{(0)}+\hbar\phi_I^{(1)})+i\hbar\gamma^\mu\nabla_\mu\phi_I^{(0)}-m_I(\phi_I^{(0)}+\hbar\phi_I^{(1)})\right]\nonumber\\
    &\quad+\mathcal{O}(\hbar^2)=\frac{(\mathfrak{m}-m_I)^2}{m_{I,II}^2}(\bar{\Phi}_I\Phi_I)_{\mathcal{O}(\hbar)}+\mathcal{O}(\hbar^2),
\end{align}
{\color{white}...........}
\begin{align}\label{App4NablaPhiIIPhiI}  \hbar\nabla^\mu\bar{\Phi}_{II}\Phi_I&=\hbar\nabla^\mu\left(\frac{e^{-\frac{i}{\hbar}\mathcal{S}_I}}{m_{I,II}}\left[(\mathfrak{m}-m_I)(\bar{\phi}_I^{(0)}+\hbar\bar{\phi}_I^{(1)})\right.\right.\nonumber\\
&\quad\left.\left.-\frac{i\hbar}{M}\left((\nabla_\mu\pi_I^\mu)-\tfrac{i}{2}\sigma^{\mu\nu}eF_{\mu\nu}+2\pi_I^\mu\nabla_{\mu}\right)\bar{\phi}_I^{(0)}\right]\right)\Phi_I\nonumber\\
&=\frac{(\mathfrak{m}-m_I)}{m_{I,II}}\left[\hbar\nabla^\mu\bar{\phi}_I^{(0)}\phi_I^{(0)}-iP^\mu_I\left(\bar{\Phi}_I\Phi_I\right)_{\mathcal{O}(\hbar)}\right]+\mathcal{O}(\hbar^2),
\end{align}
\begin{align}\label{App4NablaPhiIIPhiII}
\hbar\nabla^\mu \bar{\Phi}_{II}\Phi_{II}&=\hbar\nabla^\mu\left(\frac{e^{-\frac{i}{\hbar}\mathcal{S}_I}}{m_{I,II}}\left[(\mathfrak{m}-m_I)(\bar{\phi}_I^{(0)}+\hbar\bar{\phi}_I^{(1)})\right.\right.\nonumber\\
&\quad\left.\left.-\frac{i\hbar}{M}\left((\nabla_\mu\pi_I^\mu)-\tfrac{i}{2}\sigma^{\mu\nu}eF_{\mu\nu}+2\pi_I^\mu\nabla_{\mu}\right)\bar{\phi}_I^{(0)}\right]\right)\Phi_{II}\nonumber\\
&\!\!\!\!\!\!=\frac{(\mathfrak{m}-m_I)^2}{m_{I,II}^2}\left[\hbar\nabla^\mu\bar{\phi}_I^{(0)}\phi_I^{(0)}-iP_I^\mu\left(\bar{\Phi}_I\Phi_I\right)_{\mathcal{O}(\hbar)}\right]+\mathcal{O}(\hbar^2).
\end{align}
These four results show that, for consistency, one needs to choose and stick throughout to one of the two possible values $m_1$ and $m_2$ of $\frak{m}$. In fact, using, for example, Eq.\,(\ref{App4PhiIIPhiII}), we have on the one hand, $\bar{\Phi}_{II}\Phi_{II}=(\frak{m}-m_I)^2(\bar{\Phi}_I\Phi_I)_{\mathcal{O}(\hbar)}/m_{I,II}^2+\mathcal{O}(\hbar^2)$. On the other hand, switching the indices $I\leftrightarrow II$ in Eq.\,(\ref{App4PhiIIPhiII}), we arrive at $\bar{\Phi}_{I}\Phi_{I}=(\frak{m}-m_{II})^2(\bar{\Phi}_{II}\Phi_{II})_{\mathcal{O}(\hbar)}/m_{I,II}^2+\mathcal{O}(\hbar^2)$. These two identities would agree with each other if and only if the reduced mass $\frak{m}$ is taken to be either $m_1$ \textit{or} $m_2$ in \textit{both} identities, in which case we do indeed obtain $(\frak{m}-m_I)^2(\frak{m}-m_{II})^2/m_{I,II}^4=1$.

\end{document}